\documentclass[pre,twocolumn,showpacs,floatfix,superscriptaddress,aps]{revtex4}
\usepackage{amsmath}
\usepackage{makeidx}
\usepackage{amssymb}
\usepackage{graphicx}

\usepackage[usenames]{color}
\usepackage[normalem]{ulem}

\newcommand{\beq}{\begin{equation}}
\newcommand{\eeq}{\end{equation}} 
\newcommand{\beqa}{\begin{eqnarray}}
\newcommand{\eeqa}{\end{eqnarray}} 
\begin{document}


\title{Statics and dynamics of a self-bound matter-wave quantum ball}

\author{ S. K. Adhikari\footnote{adhikari@ift.unesp.br; URL: http://www.ift.unesp.br/users/adhikari}
} 
\affiliation{
Instituto de F\'{\i}sica Te\'orica, UNESP - Universidade Estadual Paulista, 01.140-070 S\~ao Paulo, S\~ao Paulo, Brazil
} 

\begin{abstract}

We study  the statics and dynamics  of a stable, mobile,   three-dimensional  matter-wave spherical quantum ball created in the presence of an  attractive two-body and a { {\it very small}}
repulsive three-body interaction. The quantum ball  can  propagate with a constant velocity in any direction { in free space} and its  stability under a small perturbation is established {   numerically and variationally}. In frontal head-on  and angular  collisions at large velocities two quantum balls behave like quantum solitons. Such collision is found to be quasi elastic and the quantum balls emerge after collision without any change of direction of motion and velocity and with practically no deformation  in shape.  When reflected by a hard impenetrable plane, the quantum ball bounces off like a wave obeying the law of reflection  without any change of shape or speed. However, in a collision   at small velocities two quantum balls coalesce to form a 
larger ball which we call a quantum-ball breather. We  point out the similarity and difference between the collision of two quantum and classical balls. The present study is based on an analytic  variational approximation and a full numerical solution of the mean-field Gross-Pitaevskii equation  using the parameters of $^7$Li atoms.

\end{abstract}


\maketitle

 \section{Introduction}
 
After the experimental observation of Bose-Einstein condensate (BEC) \cite{expt1,rmp1999}, there have been 
many experimental studies to investigate different  quantum phenomena in a laboratory 
previously not accessible for study in a controlled environment, such as, quantum 
phase transition \cite{qpt}, quantum collapse of matter wave under attraction \cite{bosenova}, four-wave mixing 
of matter waves \cite{4wm}, formation of vortex lattice \cite{vl}, interference of matter waves \cite{imw}, Josephson tunneling \cite{jos}, Anderson localization \cite{ander}
 etc. 
The generation and the dynamics of self-bound large   quantum waves have drawn 
much attention lately. There have been some studies of self-bound matter waves or solitons
in one (1D) \cite{rmp} or  two  (2D) \cite{santos} space dimensions.
A (self-bound) bright soliton  travels at a constant velocity  
in 1D, due to a cancellation of  nonlinear attraction and defocusing forces \cite{book,sol}. The collision between two analytic 1D
 bright solitons is always elastic \cite{book}
with the solitons emerging after collision without a change of velocity and shape.     The 1D soliton  has been observed  in nonlinear optics \cite{book} and in Bose-Einstein condensates \cite{rmp}.  
However, a two- or three-dimensional (3D)   soliton
cannot be realized for two-body contact attraction   alone
due to collapse \cite{book}. 
The  1D BEC solitons studied experimentally \cite{rmp}
 are really  quasi solitons behaving like  real solitons at large velocities. At low velocities a deformation of their shapes is expected upon collision. 

 There have been a few proposals for creating a self-bound  3D
  matter-wave state which we term a quantum ball.  Some of these proposals involve an engineering of the atomic scattering length using the Feshbach resonance technique to generate  dynamically stabilized  solitons in 2D and 3D 
  \cite{sadhan}. 
Others consider extra interactions usually neglected in dilute  BEC of alkali atoms to create a stationary localized state. In the presence of an additional nonlocal dipolar interaction a 2D 
BEC soliton can be generated \cite{santos}.  
 It has been suggested by Maucher {\it et al.}  \cite{ryd} that in the case of Rydberg atoms, off-resonant dressing to Rydberg nD states can provide a nonlocal long-range attraction which can form a quantum ball. The collapse instability can be stopped in this case by a repulsive contact interaction.  

In this paper we demonstrate that {a {\it very small} repulsive} three-body interaction in the presence of an attractive two-body contact interaction can generate a stable quantum ball. The collapse is stopped in this case 
by the repulsive three-body interaction.
 Although, some theoretical suggestions for generating a quantum ball, in the presence of  a repulsive core in the two-body atomic interaction, seem viable \cite{luca,ryd},  so far there has not been success 
in their experimental realization. 
Nevertheless, there are questions about the dynamics of a quantum ball which are very intriguing. 
The study of the dynamics of a quantum elementary  particle like an electron to determine simultaneously its position and velocity
is doomed to failure due to the Heisenberg uncertainty relation. On the other hand, for a  quantum ball the uncertainty relation is not of concern due to its large mass and it can be traced like a classical object by its position and velocity at each instant.     
As the quantum ball is self bound it can move like a classical ball obeying Newton's first law of motion. However, very little is known about the interaction dynamics of two quantum balls and that of a quantum ball with other objects and we address these questions in this paper.

We consider the mean-field Gross-Pitaevskii (GP) equation with the inclusion of a three-body interaction   for the study of statics and dynamics  of a quantum ball.
The quantum balls   are bound by an attractive two-body contact interaction in the presence of a repulsive three-body contact interaction. 
We use the realistic parameters of $^7$Li atoms, in this study, with a negative scattering length 
corresponding  to two-body attraction \cite{rmp1999}. {The effect of atom loss due to three-body recombination is included in the study of dynamics.}
 {  The three body loss rate is not accurately known for
this system \cite{lossrate} for the parameter domain used in this study.    We have chosen three-body loss rates that ensure our
system does not decay significantly during our dynamical simulations.}
It is expected that the dynamics of quantum balls will be independent of the details of the mechanism responsible for self binding and we do not believe that the results obtained  here are so peculiar as to have no general validity.  {In fact, a preliminary study revealed similar dynamics for quantum balls made of dipolar atoms \cite{unpub}.}
 A stationary quantum ball  can be formed for the two-body attraction above a critical value  in the 3D GP equation for any finite three-body repulsion. 
The statical properties  of the quantum ball are  studied using a   variational analysis
and a   numerical solution  of the   3D GP equation.   The variational and numerical 
results are found to be in good agreement with each other. The stability of the quantum ball  is established numerically 
under a small perturbation introduced by changing the  three-body interaction by a small amount, while the quantum ball  is found to execute sustained breathing oscillation.

A quantum ball   can move freely without deformation along any direction  with a constant velocity. 
We   study the frontal and angular  collisions between two quantum balls. Only the collision between two integrable 1D solitons is truly elastic \cite{book}. As the dimensionality of the  
soliton is increased such collision is expected  to become inelastic with loss of energy in 2D and 3D.  
In the present numerical simulation of frontal collision between two  quantum balls, 
at sufficiently large velocities   the collision is found to be quasi elastic when the two quantum balls emerge after collision with practically no deformation and without any change of velocities. Unlike classical balls, obeying 
the Newton's laws, there is no change in the directions of motion of the quantum balls after collision.  However, upon impact with a rigid impenetrable plane the quantum ball 
bounces like a classical elastic ball obeying the usual laws of reflection. 
 At small   velocities 
the collision   between two quantum balls 
is inelastic and  the quantum balls   form a single bound entity in an excited state executing breathing oscillation, which we call a
quantum-ball breather.

We present the 3D GP equation used in this study  in Sec. 
\ref{II} and a   variational analysis of the same for an analytic understanding of the formation of the quantum ball. 
In Sec. \ref{III} we present the numerical results for stationary profiles of a quantum ball.  We present numerical tests of stability of a quantum ball  under a small perturbation. 
The quasi-elastic nature of collision 
of two quantum balls  at large velocities and the formation of a quantum-ball breather  at small velocities 
 are  demonstrated by real-time simulation. 
We end with a summary of our findings in Sec. \ref{IV}.

\section{Mean-field model}

\label{II}

We consider a quantum ball in the presence of a
three-body interaction 
and  the mean-field model 
appropriate for this study.  
{The  mean-field} GP  equation  for  $N$ atoms of mass $m$ is    \cite{rmp1999}
\begin{align}
 i \hbar \frac{\partial \phi({\bf r},t)}{\partial t}=&
{\Big [} -\frac{\hbar^2}{2m}\nabla^2- \frac{4\pi \hbar^2|a|N}{m} \vert \phi \vert^2
\nonumber \\ &
+ \frac{\hbar N^2 K_3}{2} \vert \phi \vert^4
{\Big ]}  \phi({\bf r},t),
\label{eq1}
\end{align}
where $a$ is the scattering length, 
and $K_3$ is the three-body interaction term. The negative scattering length $a$ represents two-body attraction and the positive  $K_3$ to three-body repulsion.
 
For an analytic understanding of the formation of a quantum ball  
{convenient } variational approximation of Eq. (\ref{eq1}) can be obtained with
the following Gaussian ansatz for the time-independent stationary  wave function \cite{pg}
\begin{eqnarray}\label{eq3}
 \phi({\bf r})=\frac{\pi^{-3/4}}{w^{3/2}}\exp\left[-\frac{r^2}{2w^2}\right],
\end{eqnarray}
where $r^2=x^2+y^2+z^2$,  $w$ is the width.
 The energy density corresponding to Eq. (\ref{eq1}) is given by
\begin{eqnarray}\label{eq4}
{\cal E}({\bf r})=
\frac{\hbar^2|\nabla \phi({\bf r}) |^2}{2m}-\frac{2\pi N |a|\hbar^2 |\phi({\bf r})|^4}{m}
+\frac{\hbar N^2K_3| \phi({\bf r})|^6}{6}.
\end{eqnarray}
Consequently, the total energy per atom  $E\equiv  \int {\cal E}({\bf r}) d{\bf r}$ becomes
\begin{eqnarray}\label{eq5}
E= \frac{\hbar^2}{m}\frac{3}{4w^2} -\frac{4\pi N |a|\hbar^2}{m}\frac{\pi^{-3/2}}{4\sqrt 2  w^3}
+\frac{\hbar N^2 K_3}{2}\frac{\pi^{-3}}{9\sqrt 3  w^6}.
\end{eqnarray}
The width $w$ of a {stationary quantum ball with negative energy corresponds to a global} minimum of energy $E$:
\begin{eqnarray}\label{eq6}
 \frac{1}{w^3} -\frac{4\pi N |a|}{  (2\pi)^{3/2}w^4}+\frac{m N^2 K_3}{2\hbar}\frac{4\pi^{-3}}{9\sqrt 3 w^7}=0.
\end{eqnarray}
Without the quintic term ($K_3=0$) the quantum ball of width $w=4\pi N|a|/(2\pi)^{3/2}$  is tantamount  to an unstable Towne's soliton \cite{townes}. 
  For stability  a 
non-zero quintic term ($K_3> 0$) is necessary.

\begin{figure}[!t]

\begin{center}
\includegraphics[width=.9\linewidth,clip]{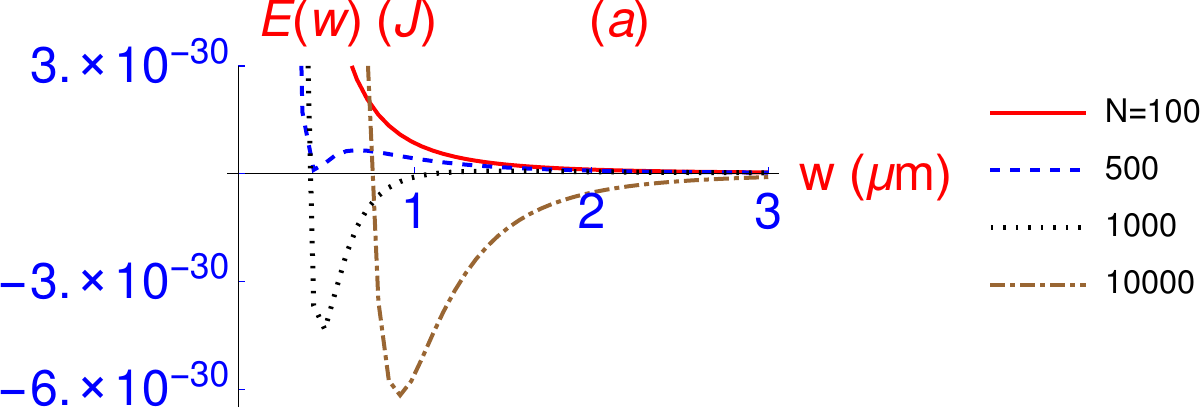} 
\includegraphics[width=.8\linewidth,clip]{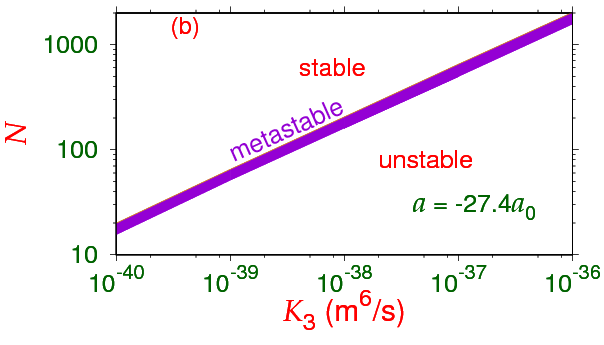}
\includegraphics[width=.8\linewidth,clip]{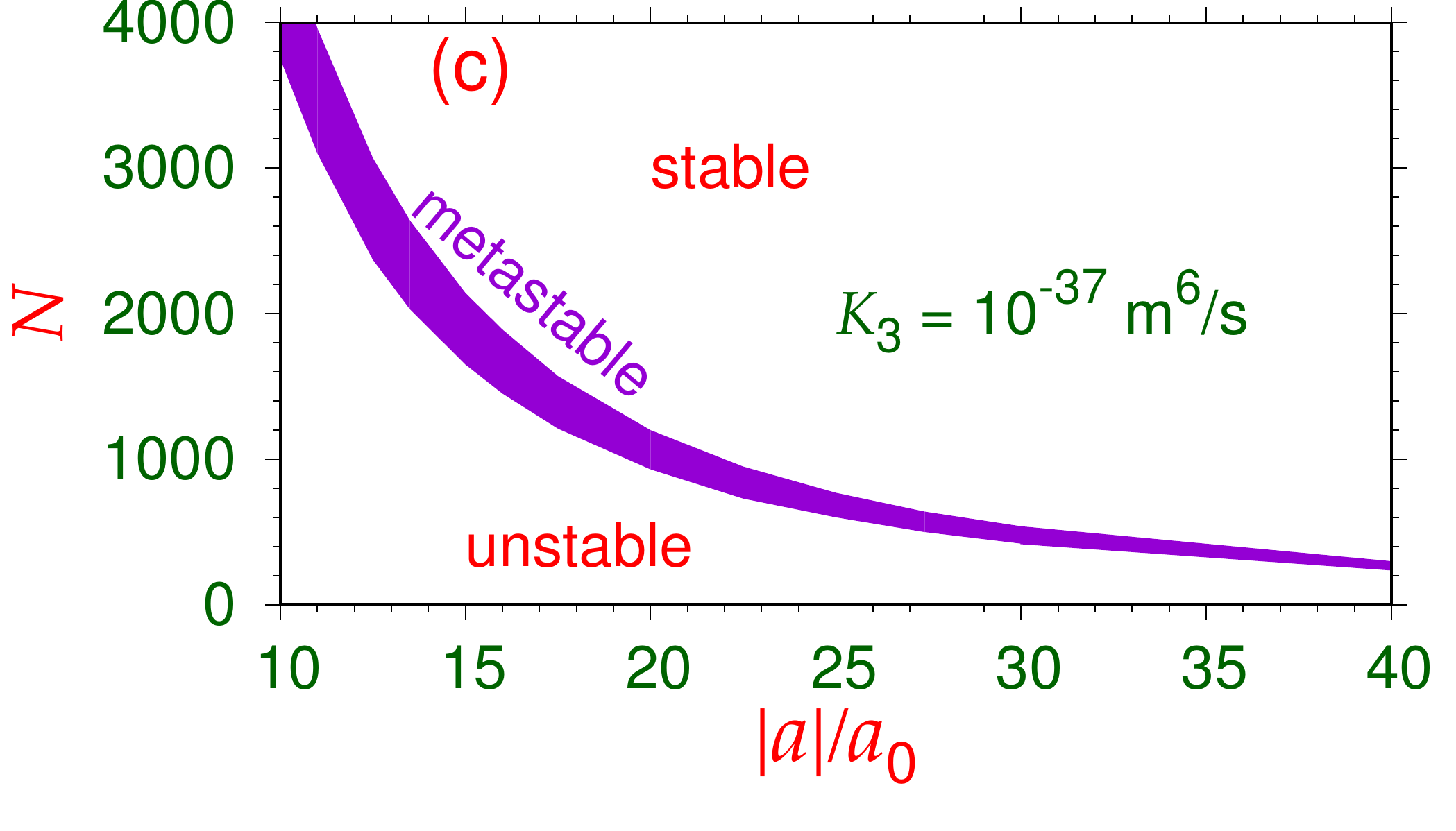} 
\caption{ (Color online) (a) The variational energy  $E$  versus $w$ 
in $\mu$m
  for different $N$ from Eq. (\ref{eq6}) for $a=-27.4a_0$ and $K_3=10^{-37}$ m$^6$/s.   Variational (b)  $N-K_3$ and
(c) $N-a/a_0$ phase plots  for $a=-27.4a_0$ and $K_3=10^{-37}$ m$^6$/s,  respectively, illustrating the regions of formation of a stable and metastable quantum ball   obtained from Eq. (\ref{eq6}). } 
\label{fig1} \end{center}

\end{figure}

\section{Numerical Results}

\label{III}

In the numerical calculation, we use the parameters of $^7$Li atoms, e.g., 
$a\approx -27.4 a_0$ \cite{rmp1999,dal} and $m= 7$ amu, where $a_0$ is the Bohr radius. 
{ Unlike the 1D case, the 3D GP equation (\ref{eq1}) does
not have analytic solution and different numerical methods, such as split-step Crank-Nicolson \cite{CPC}  and Fourier
spectral \cite{spec}  methods, are used for its solution.}
We solve  the 3D GP equation (\ref{eq1}) numerically
by the split-step 
Crank-Nicolson method using both real- and imaginary-time propagation
  in Cartesian coordinates  
using a space   step of  {$ 0.025$ $\mu$m
and a time step of  $ 0.00002$ ms in all calculations  \cite{CPC}.  All imaginary-time simulations were performed in a box of size $240\times 240 \times 240$ unless otherwise stated.}
Imaginary-time simulation is employed to get  the lowest-energy bound state  of a quantum ball,
while the real-time simulation is to be used to study the dynamics  using the initial profile obtained in the  imaginary-time  propagation \cite{CPC}.
There are different C and FORTRAN programs for solving the GP equation \cite{CPC,CPC1}
and one should use the appropriate one. 
In the imaginary-time propagation the initial  state was taken as  in Eq. (\ref{eq1})
 and the width $w$ set equal to  the variational solution obtained by solving Eq. (\ref{eq6}). 
The convergence will be quick if the guess for the width $w$ is close to the final width.

\begin{figure}[!t]

\begin{center}

\includegraphics[width=\linewidth,clip]{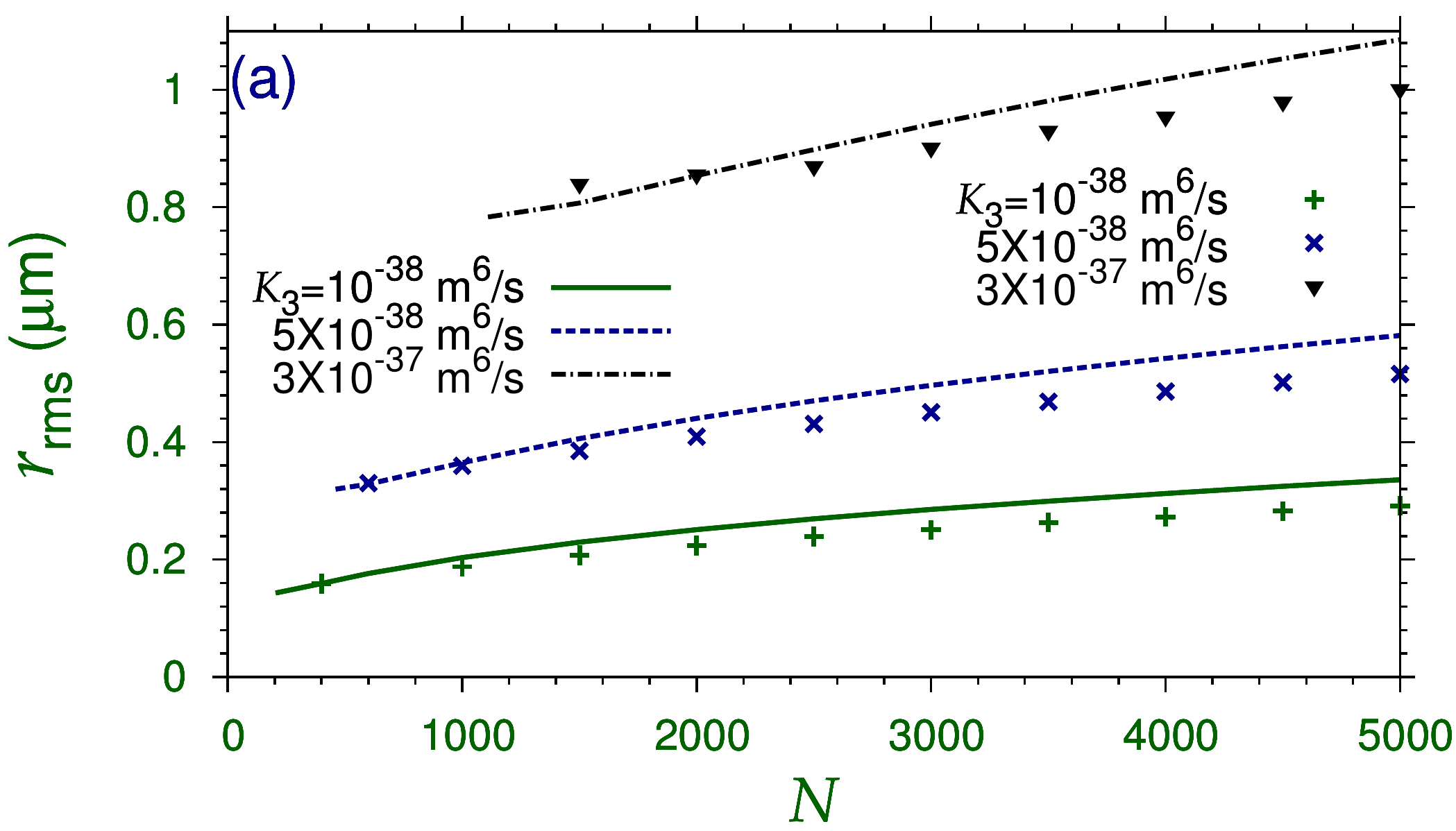}
\includegraphics[width=\linewidth,clip]{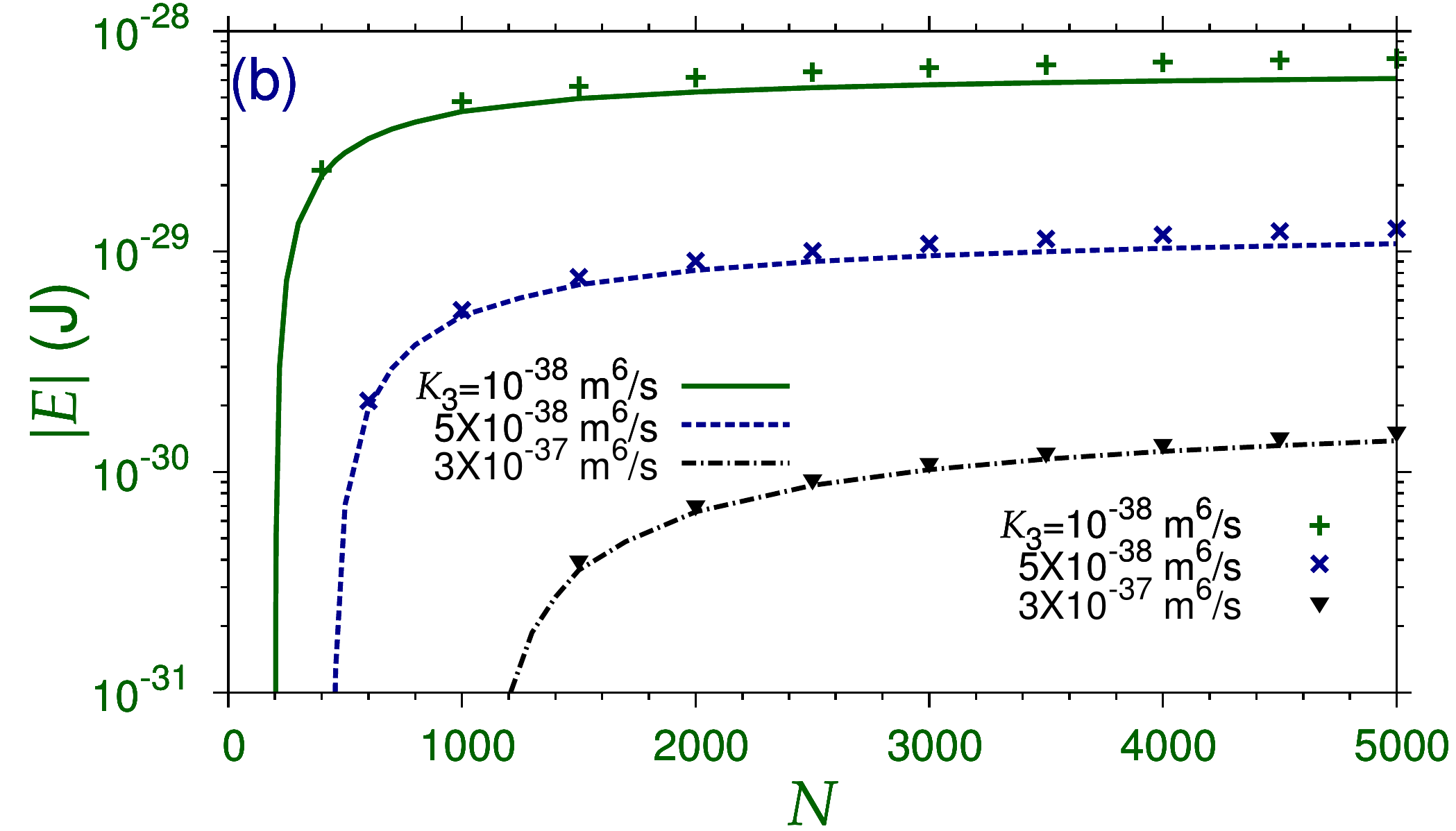}
\includegraphics[width=\linewidth,clip]{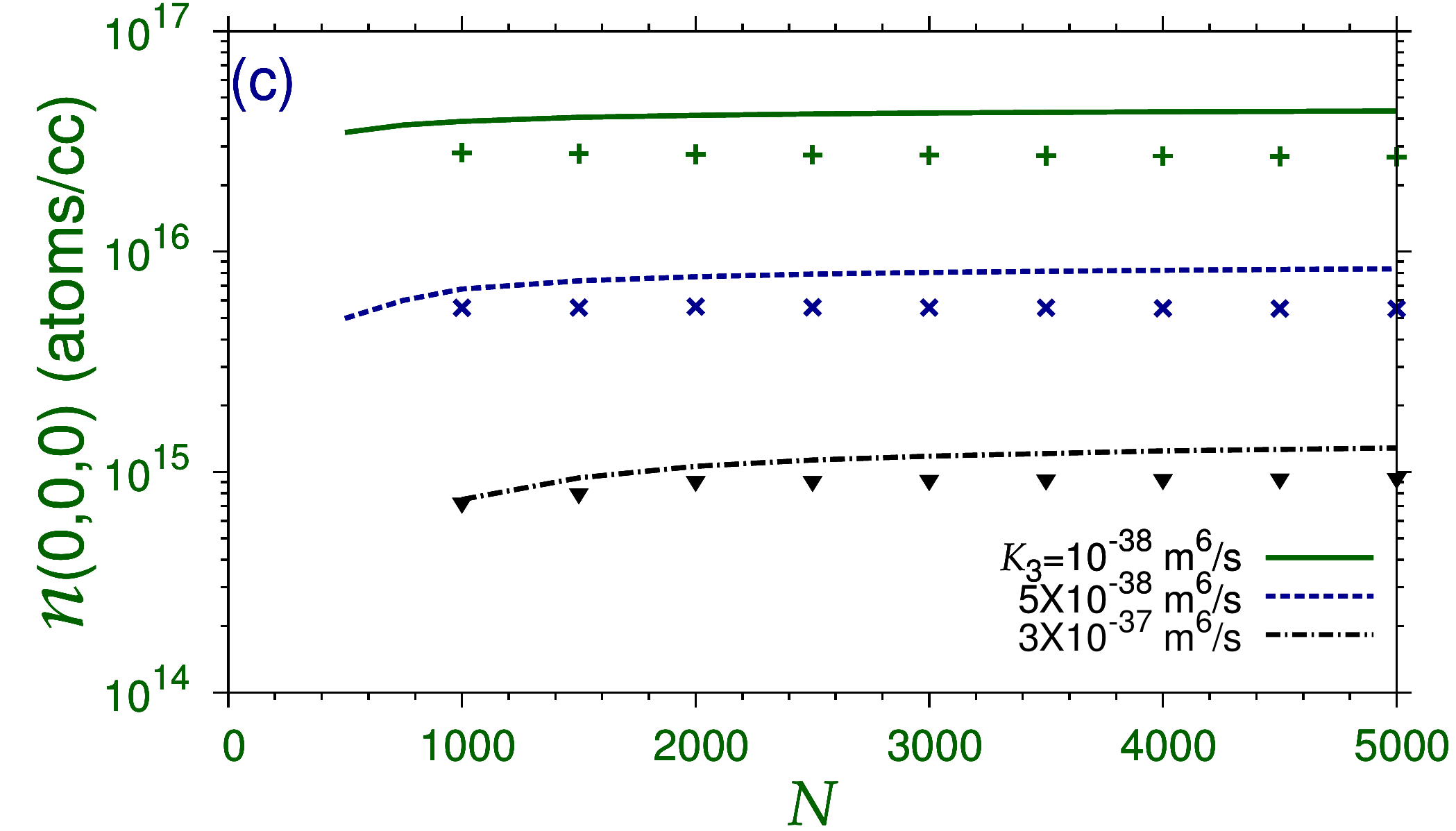}
 
\caption{ (Color online) 
  Variational (line) and numerical (points) (a)  rms radius  $r_{\mathrm{rms}}$,  (b) energy $|E|$  and (c) peak atom density {$n(0,0,0)\equiv N|\phi(0,0,0)|^2$}
versus  the number of $^7$Li atoms  $N$ in a  quantum ball 
for  three
different three-body interactions $K_3= 10^{-38}$ m$^6$/s, $= 5\times 10^{-38}$ m$^6$/s,
$= 3\times 10^{-37}$ m$^6$/s.  
}\label{fig2} \end{center}

\end{figure}

The variational width of a stationary quantum ball can be obtained from a solution of Eq. 
(\ref{eq6}).
{If a minimum of energy (\ref{eq5}) exists, it can be either  a global minimum with negative energy corresponding to a stable state or a local minimum of positive energy corresponding to a metastable state.   The energy $E\to \infty$ at $w=0$ { even for a very small non-zero $K_3$}, and $E=0$ as $w\to \infty$. { Hence collapse is not allowed in the presence of a very small three-body repulsion.}   
For certain values of the parameters there is a negative energy region between these 
two limiting values where the global minimum corresponding to a stable stationary state is located. For other sets of parameters, the energy changes monotonically between the above two limiting values without ever becoming negative or may have a local minimum with positive energy corresponding to a metastable state.    This is illustrated in Fig. \ref{fig1}(a) for $^7$Li atoms by a  plot of $E$ versus $w$ for different $N$ from Eq. (\ref{eq5}) for $a=-27.4a_0$ and $K_3=10^{-37}$ m$^6$/s.   For $N=100$ there is no minimum of $E$ and there cannot be a quantum ball. For $N=500$ there is a minimum at positive energy corresponding to a metastable state. Finally,  for $N=1000$ and 10000 there are minima at negative energies corresponding  to stable states.  The parameter domains for the formation of stable and metastable states are shown in $N-K_3$ and $N-|a|/a_0$ phase plots for $a=-27.4a_0$ and 
$K_3=10^{-37}$ m$^6$/s, respectively, in Figs. \ref{fig1}(b) and (c).  } An interesting scaling relation $N \sim \sqrt{K_3}$ is noted in Fig. \ref{fig1}(b).
Although there is a lower limit on the number of atoms $N$ for the formation of a stable quantum ball, viz. Figs.  \ref{fig1}(b) and (c), there is no 
upper limit on $N$.
 In the following we will only be concerned with the global minimum with  negative energy corresponding to a stable stationary state.

Next we compare  in Fig. \ref{fig2}(a) the numerical and variational 
root-mean-square (rms) radius $r_{\mathrm{rms}}$ of a $^7$Li  quantum ball
versus number of atoms $N$ for three different values of the three-body term:
$K_3= 10^{-38}$ m$^6$/s, $= 5\times 10^{-38}$ m$^6$/s,
$= 3\times 10^{-37}$ m$^6$/s. 
The variational result for the rms radius is given by:  $r_{\mathrm{rms}}= \sqrt{3/2}{w}$,
where $w$ is the equilibrium variational width.
For small $N$, the quantum balls are well localized with small size and the 
agreement between numerical and variational results is better.  In Fig. \ref{fig2}(b) we 
show the numerical and variational   energies $|E|$ of a  quantum ball 
versus $N$ for different $K_3$. The energy of a bound quantum ball is negative in 
all cases and its absolute value is plotted.  {In Fig. \ref{fig2}(c) the   numerical and variational  peak atom density 
$n\equiv N |\phi(0,0,0)|^2$ 
of the quantum ball  versus $N$  is illustrated for different  $K_3$. 
}

\begin{figure}[!t]

\begin{center}
\includegraphics[width=\linewidth,clip]{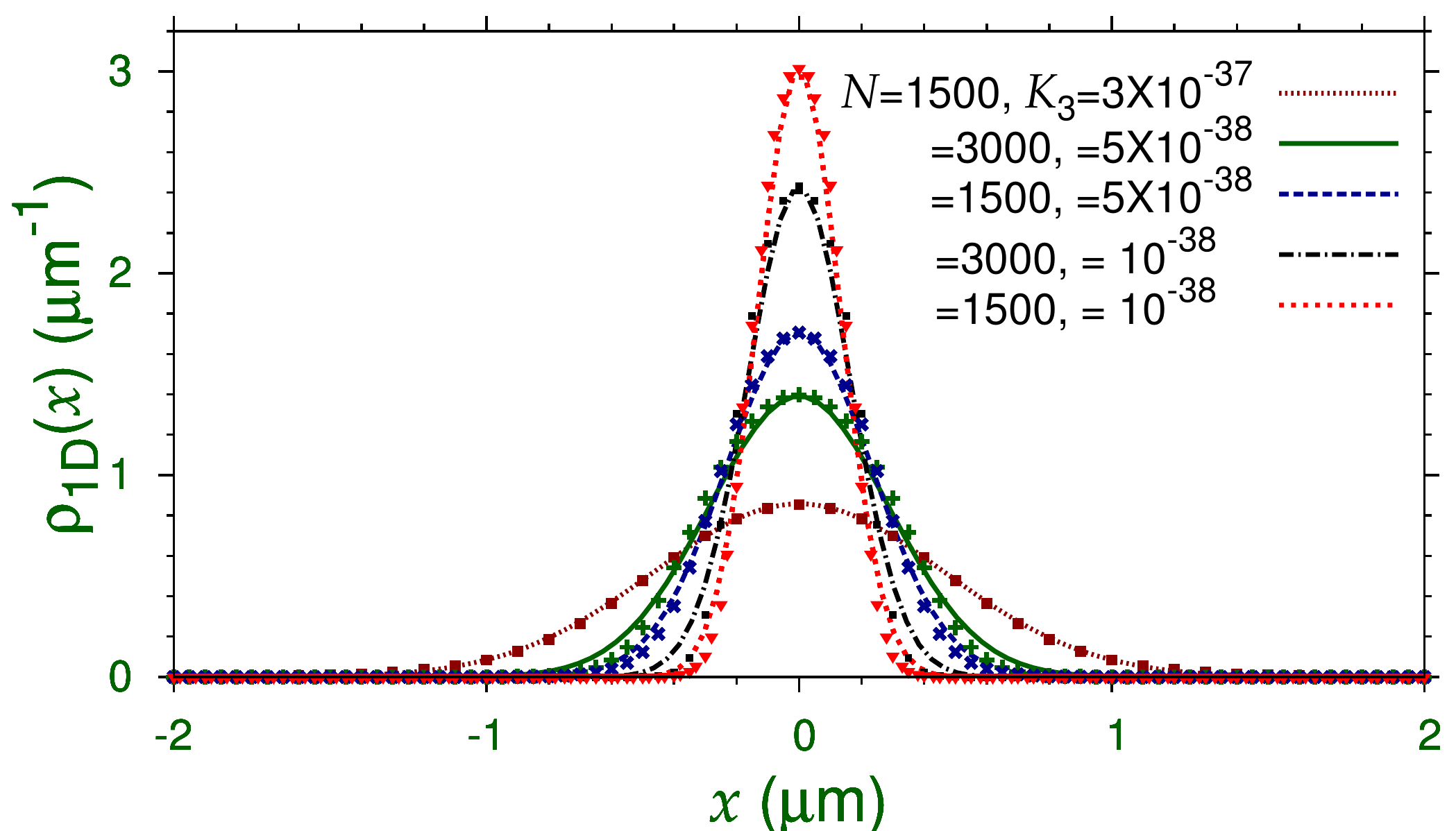}

\caption{(Color online) Numerical (chain of symbols) and  variational (line) reduced 1D density $\rho_{1D}(x)$  of a  $^7$Li quantum ball for different  $N$ and $K_3$.
}\label{fig3} 

\end{center}

\end{figure}

To study the density distribution of a $^7$Li   quantum ball we calculate the 
reduced 1D density defined by 
\begin{align}
\rho_{\mathrm{1D}}(x) = \int dz dy |\phi({\bf r})|^2.
\end{align}
In Fig. \ref{fig3} we plot this reduced 1D density as obtained from variational and numerical 
calculations for different $N$ and $K_3$. 
For a fixed  three-body term $K_3$, 
  the  quantum ball is more compact with the decrease
of number of atoms $N$.    
For a fixed  number of atoms $N$, the  quantum ball is more compact 
for a small three-body term $K_3$. 
The agreement between the two densities is better for the compact quantum balls of smaller 
size as in Fig. \ref{fig2}.


 



\begin{figure}[!t]

\begin{center}
\includegraphics[width=\linewidth,clip]{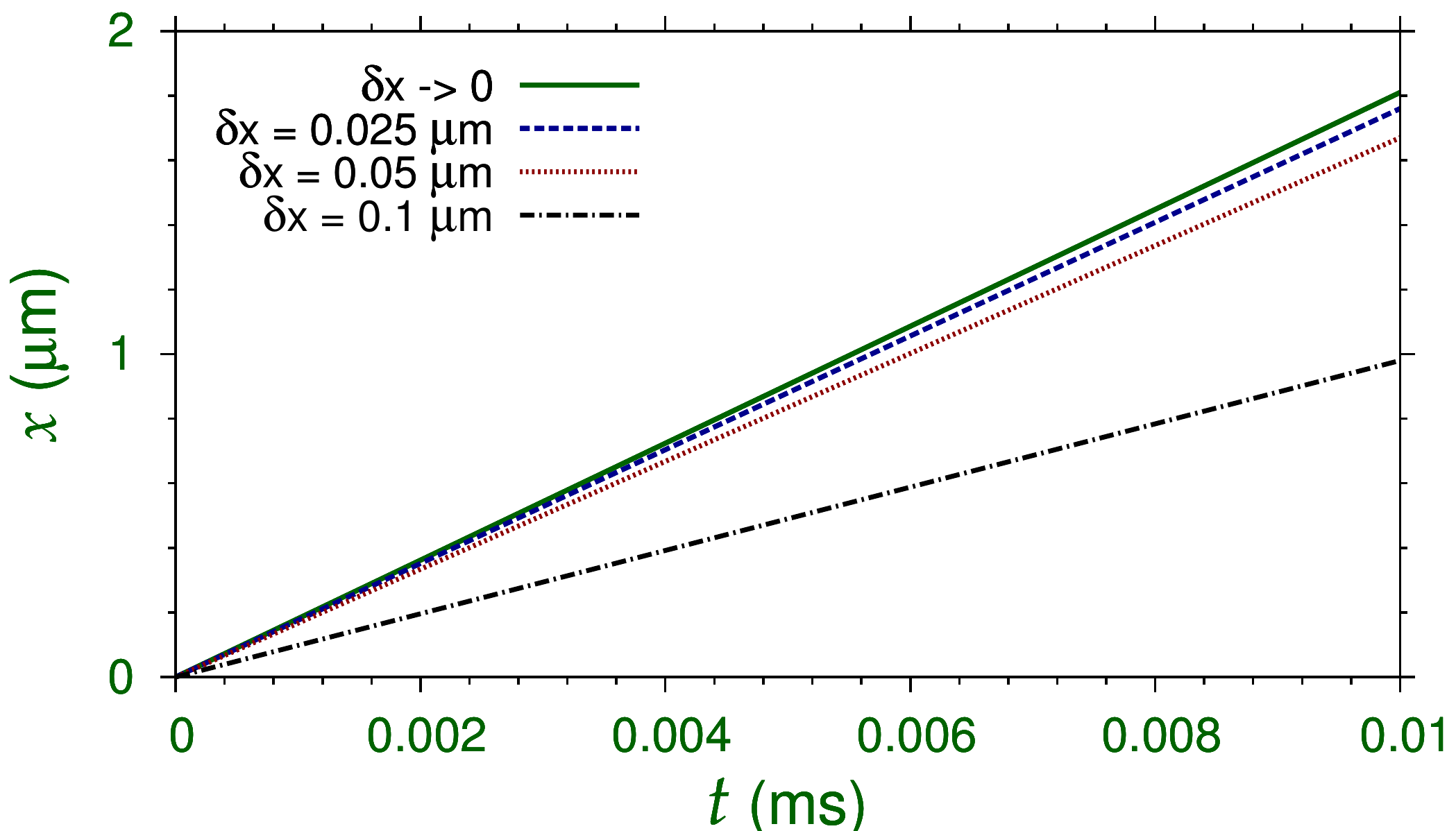}

\caption{(Color online)The displacement versus time for the quantum ball of $N=1500$ and $K_3=3\times 10^{-37}$ m$^6$/s moving with velocity $v=18.16$ cm/s calculated with different space steps $\delta x$.
}\label{Fig5} 

\end{center}

\end{figure}

{ 
We performed  numerical tests of stability of the  $^7$Li quantum balls 
under a small perturbation  (details not reported here).  We considered  a quantum-ball  wave function  as calculated by imaginary-time propagation  
 and performed   real-time propagation  with the imaginary-time wave function 
under a small perturbation introduced at $t=0$ upon changing the scattering length  by less than $1 \%$. 
 After 
this sudden 
perturbation   the  quantum ball starts a  breathing  oscillation.  
The  continued oscillation of the quantum ball over a long interval of time  establishes its stability.     }

To study the dynamics of the quantum balls, we need to set these in motion. This can be achieved by multiplying the 
 imaginary-time wave function  by a phase  $\exp(\pm i p x/\hbar)$ with $p/\hbar\equiv mv/\hbar $,  
where $p$ and $v$ are momentum and velocity,
  and   real-time simulation is then performed using these wave functions for the study of moving quantum balls with velocity $v$.   { However, to achieve the desired velocity numerically an accurate wave function calculated over a large 
space domain and small space and time steps are needed; otherwise     the numerically generated velocity is always less than the expected velocity $v$, except when $v$ is very small. This is illustrated in Fig. \ref{Fig5}  where we plot the displacement versus time for the moving quantum ball, along the 
$x$ axis,  of $N=1500$ and $K_3=3\times 10^{-37}$ m$^6$/s after multiplying its wave function by $\exp(i20 x)$, so that $p/\hbar\equiv mv/\hbar =20$ $\mu$m$^{-1}=20 \times 10^6$  m$^{-1}$ leading to $v=18.16$ cm/s. The numerically generated dynamics was obtained with 
 space steps $\delta x= 0.1$ $\mu$m, 0.05 $\mu$m, and 0.025 $\mu$m and illustrated in Fig. \ref{Fig5}. The ideal theoretical result for $\delta x \to 0 $ is also shown.  The result for $\delta x=0.025$ $\mu$m is satisfactory and this value of space step has been used in all calculations. }

 \begin{figure}[!t]

\begin{center}
\includegraphics[trim = 0mm 0mm 0mm 0mm, clip,width=.32\linewidth]{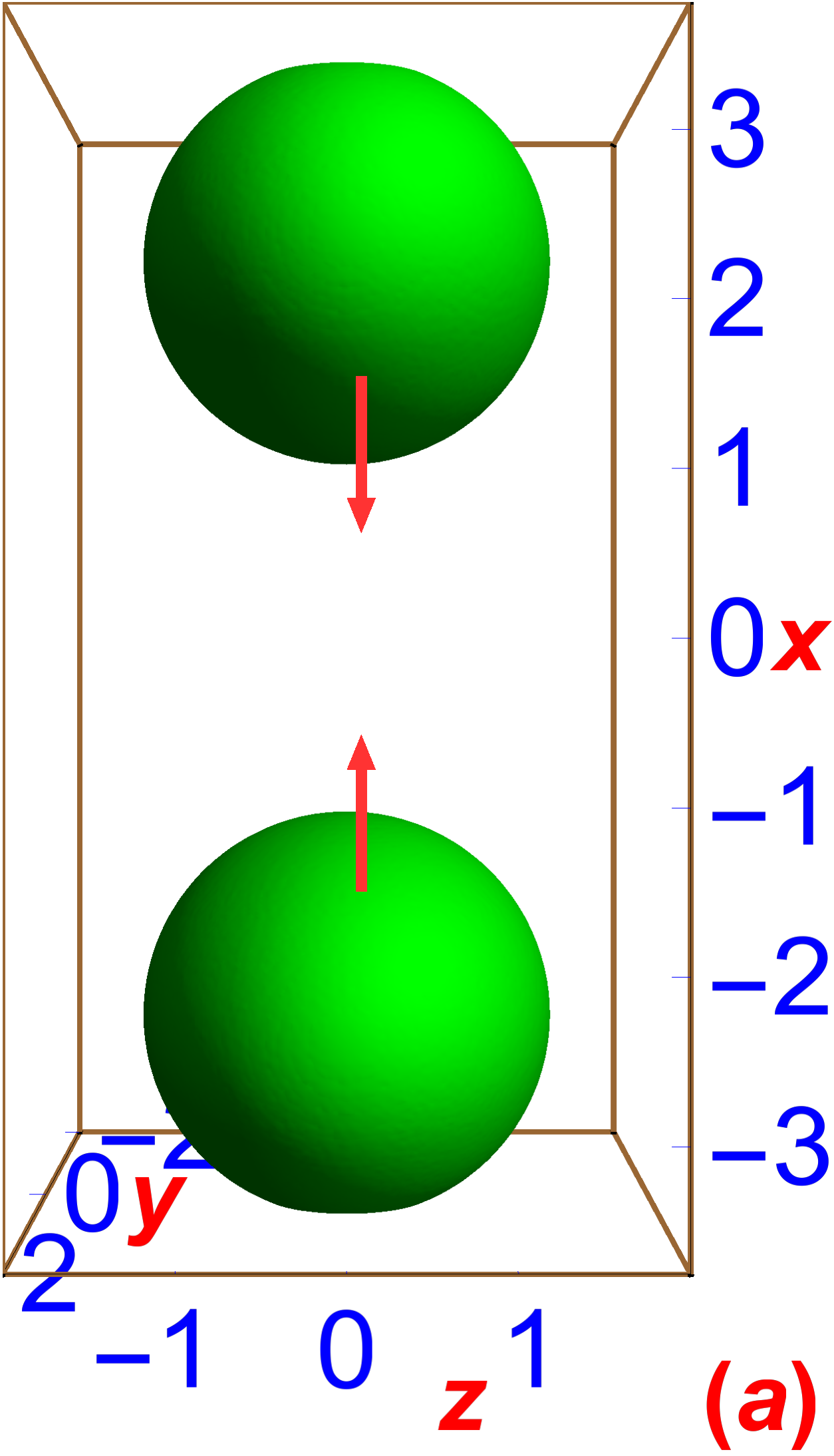}
 \includegraphics[trim = 0mm 2mm 0mm 2mm, clip,width=.32\linewidth]{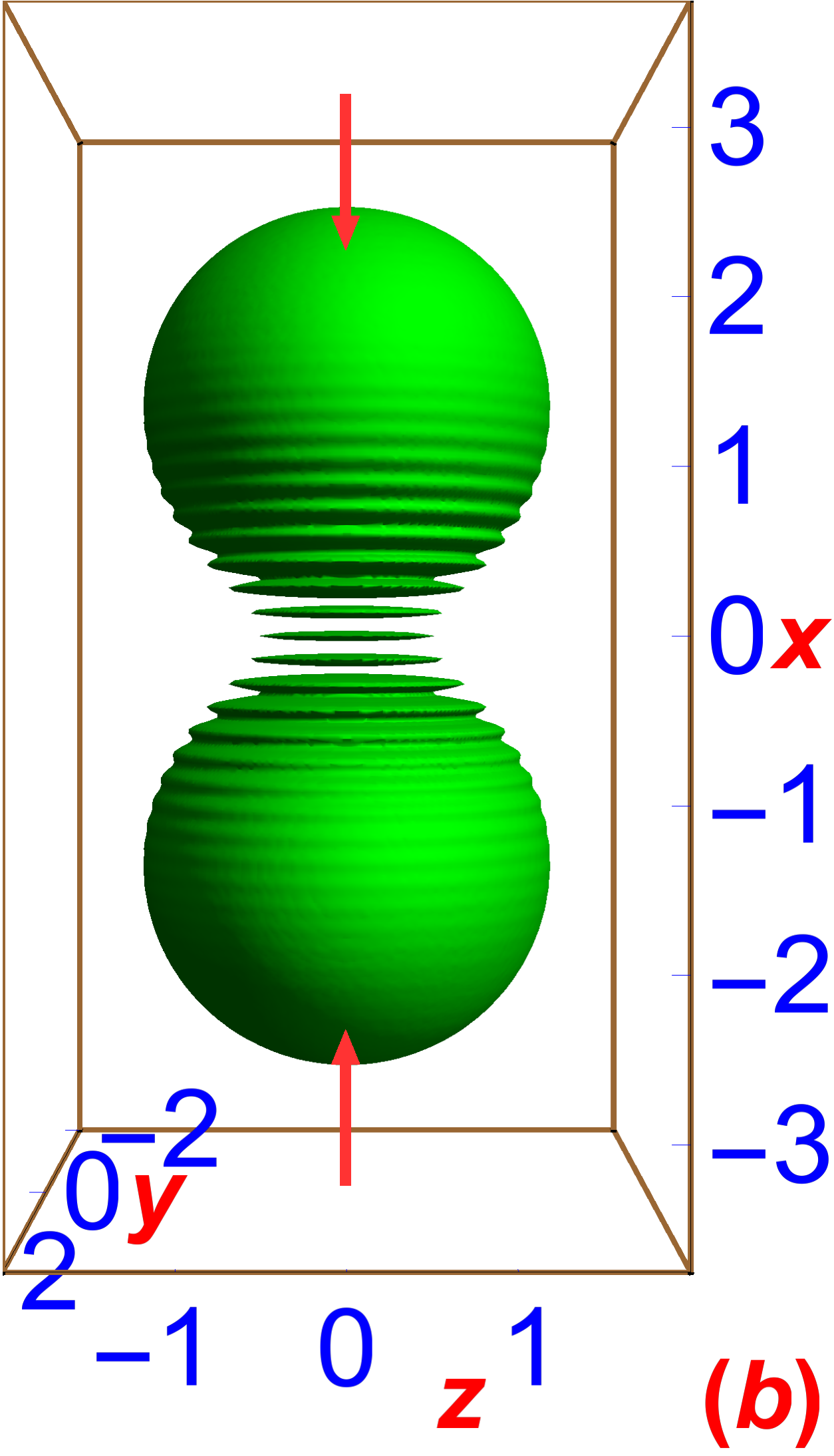} 
\includegraphics[trim = 0mm 0mm 0mm 0mm, clip,width=.32\linewidth]{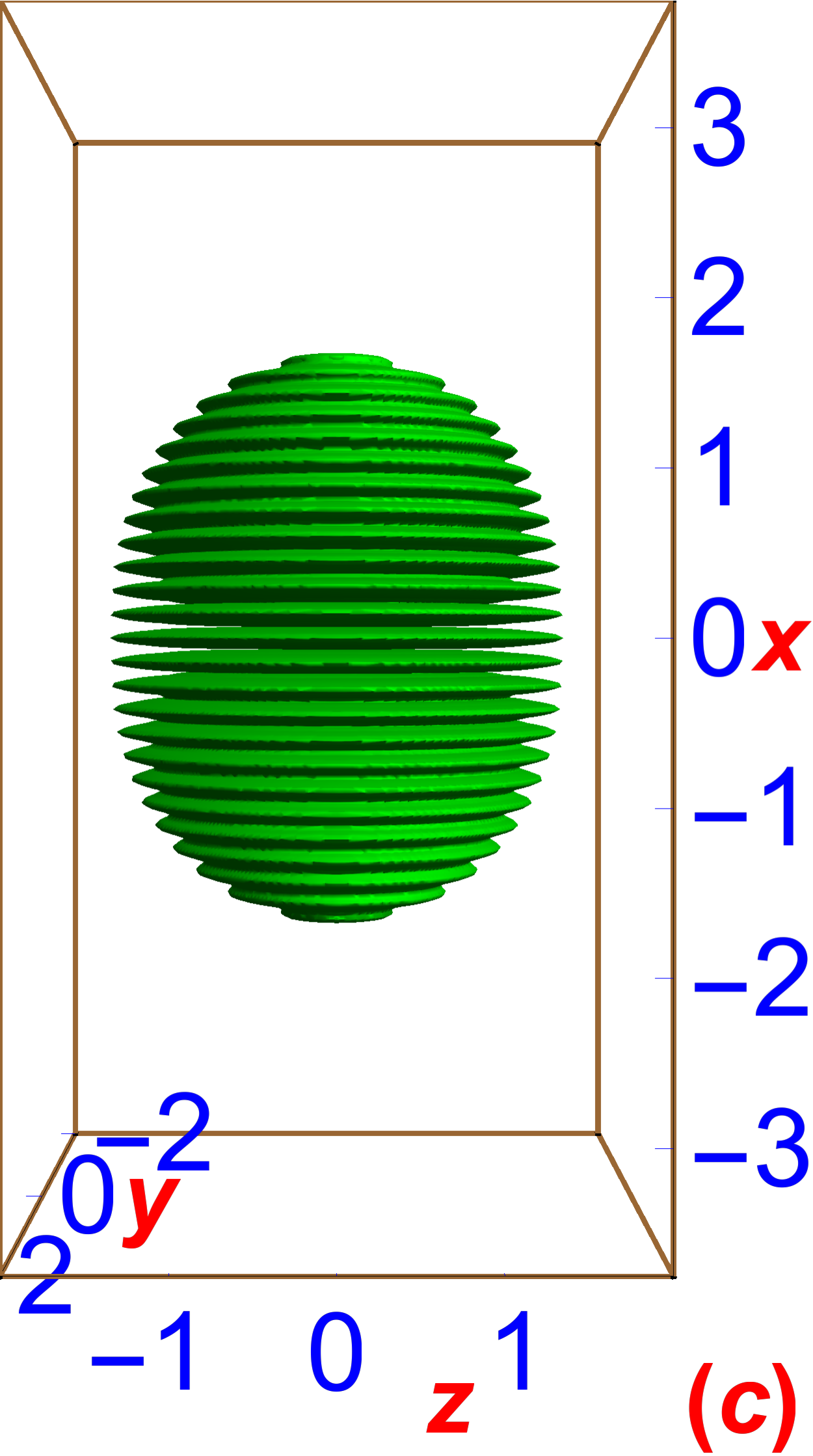}
 \includegraphics[trim = 0mm 0mm 0mm 0mm, clip,width=.32\linewidth]{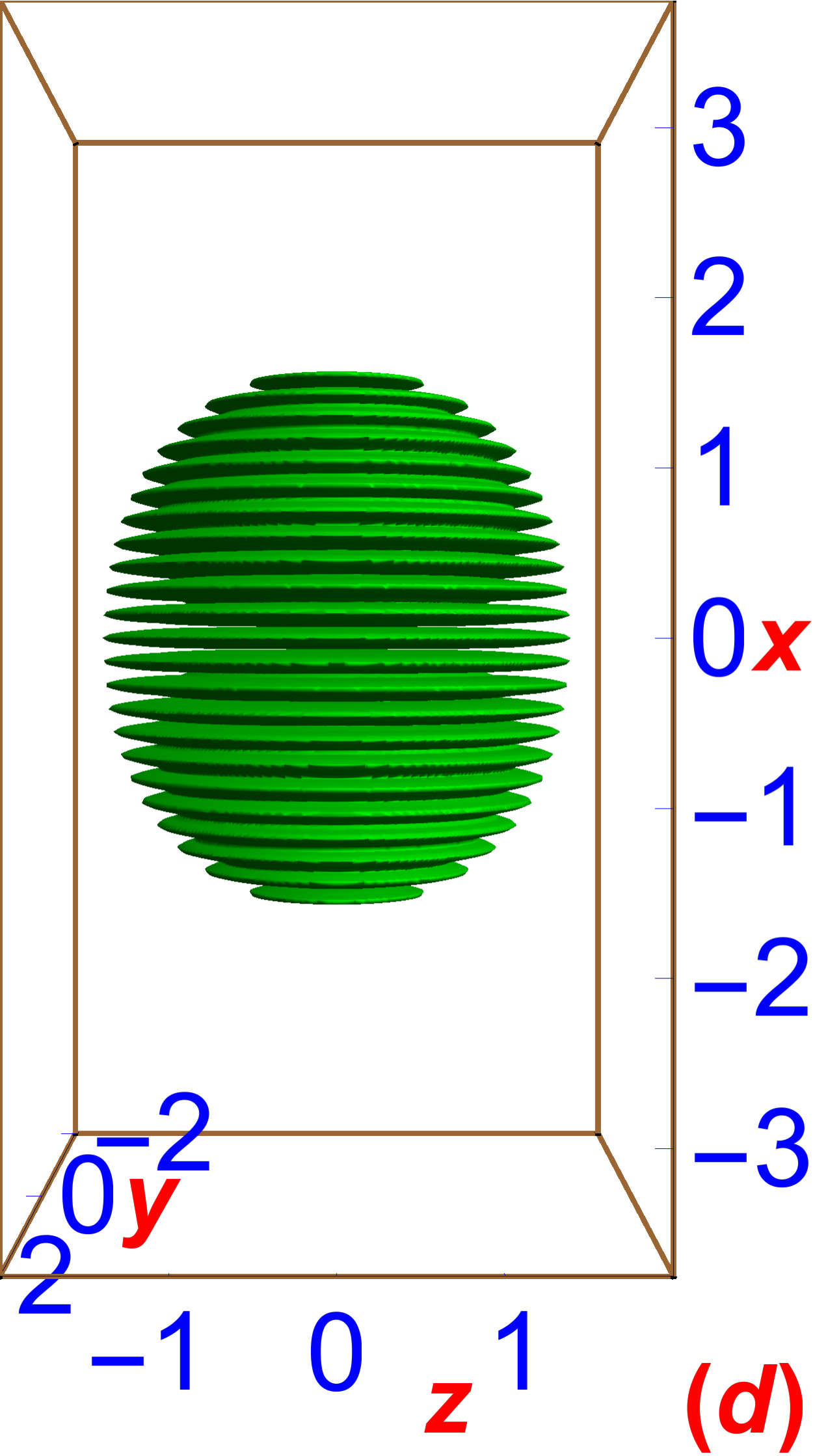}
 \includegraphics[trim = 0mm 2mm 0mm 2mm, clip,width=.32\linewidth]{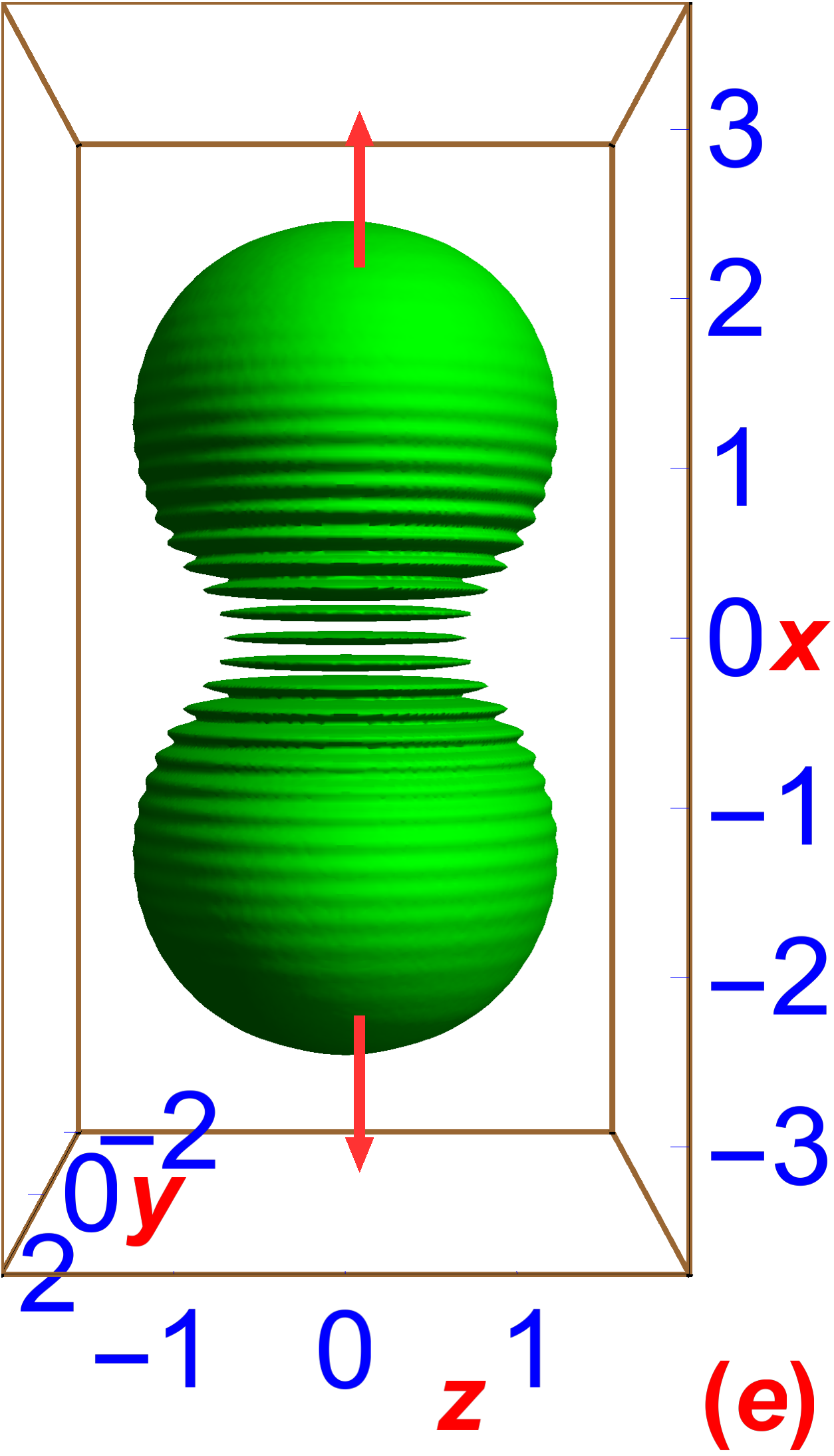} 
\includegraphics[trim = 0mm 0mm 0mm 0mm, clip,width=.32\linewidth]{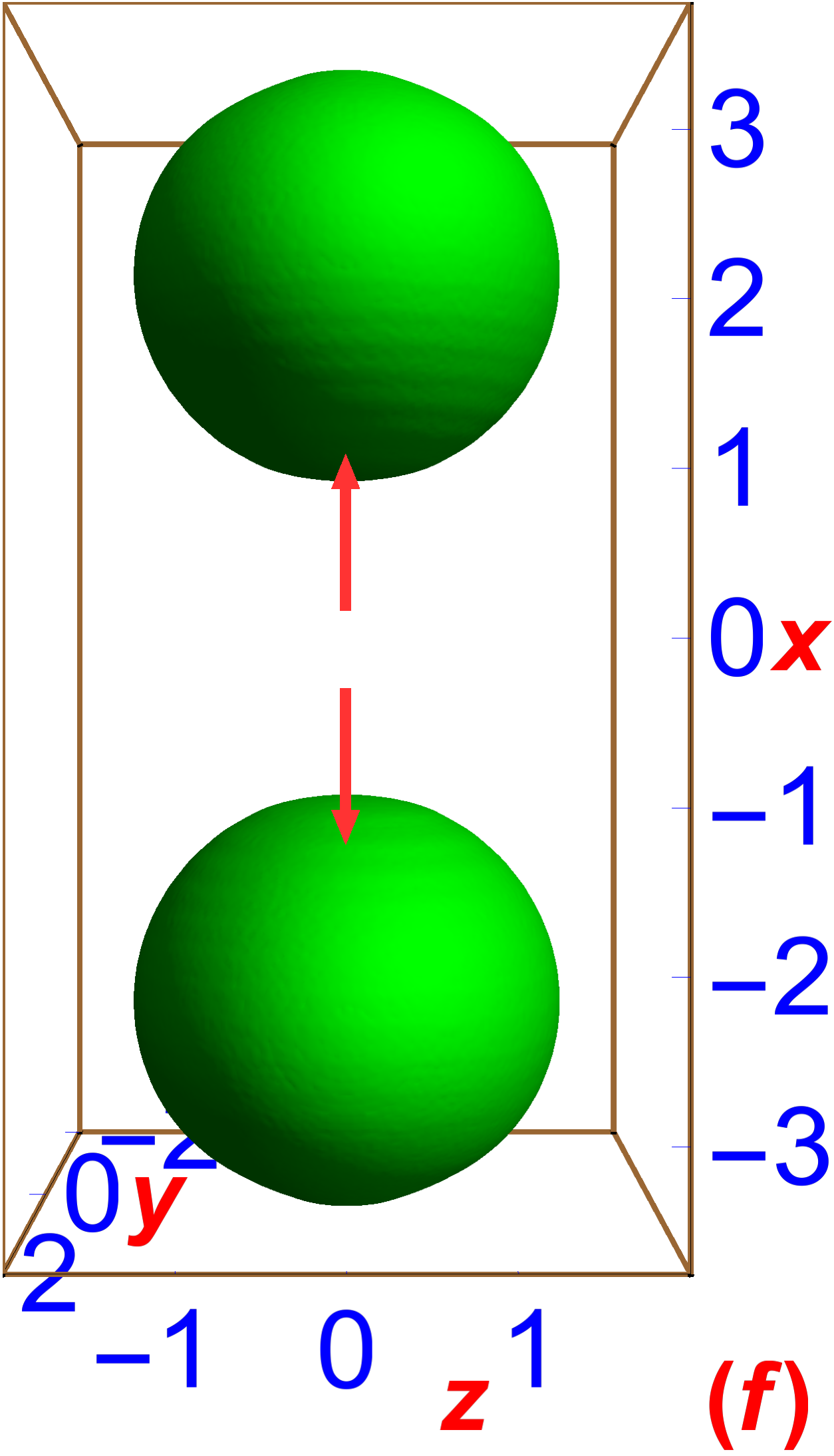}
\caption{ (Color online) Collision dynamics  of two $^7$Li quantum balls, with $N=1500, K_3=3\times 10^{-37}(1-i)$ m$^6$/s each, placed at $x=\pm 2.5$ $\mu$m at $t=0$ ms and set into motion in opposite directions along the $x$ axis with   velocity  18.16 cm/s, so as to collide at $x=0$, illustrated by   isodensity contours at times  
(a) $t=0$, (b) = 0.0057 ms, (c) = 0.0114 ms, (d) = 0.017 ms, (e) =  0.0228 ms, (f) 
= 0.0285 ms.  The density on the contour is 10$^{10}$ atoms/cm$^3$ and unit of length is $\mu$m. The directions of motion  of the  quantum balls are shown by arrows.}
\label{fig5} \end{center}

\end{figure}

 \begin{figure}[!t]

\begin{center}
\includegraphics[trim = 0mm 0mm 0mm 0mm, clip,width=.32\linewidth]{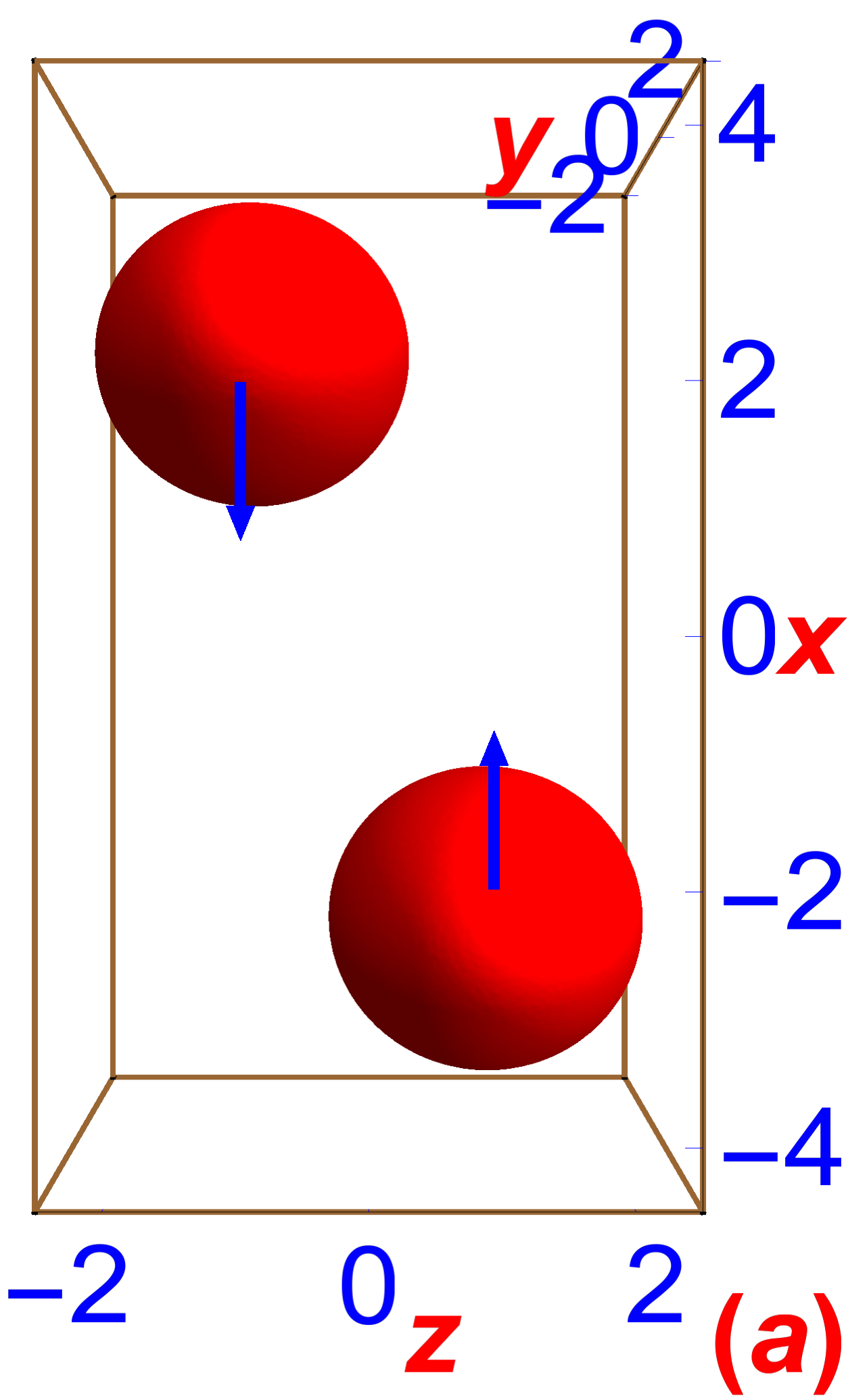}
 \includegraphics[trim = 0mm 0mm 0mm 0mm, clip,width=.32\linewidth]{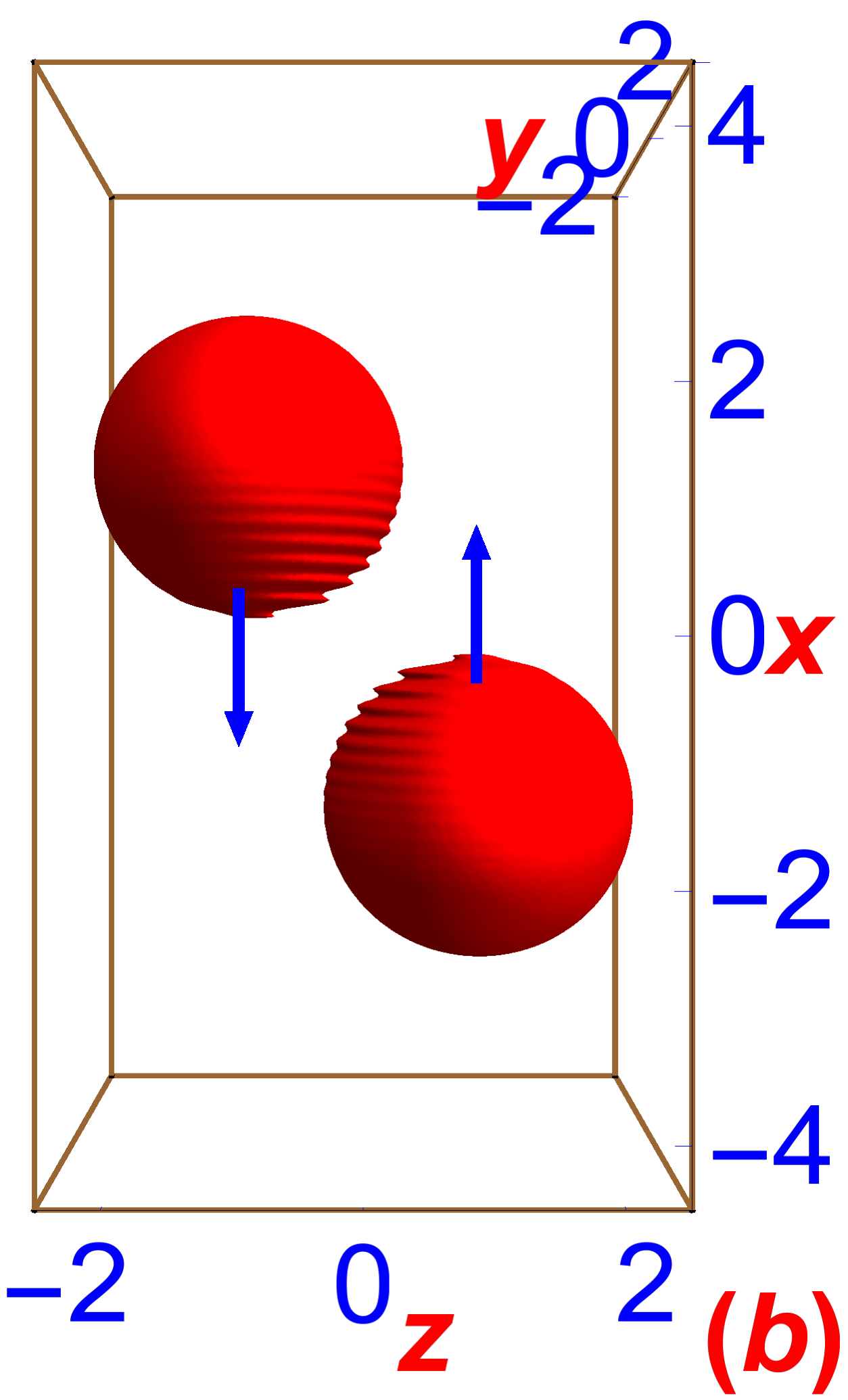} 
\includegraphics[trim = 0mm 0mm 0mm 0mm, clip,width=.32\linewidth]{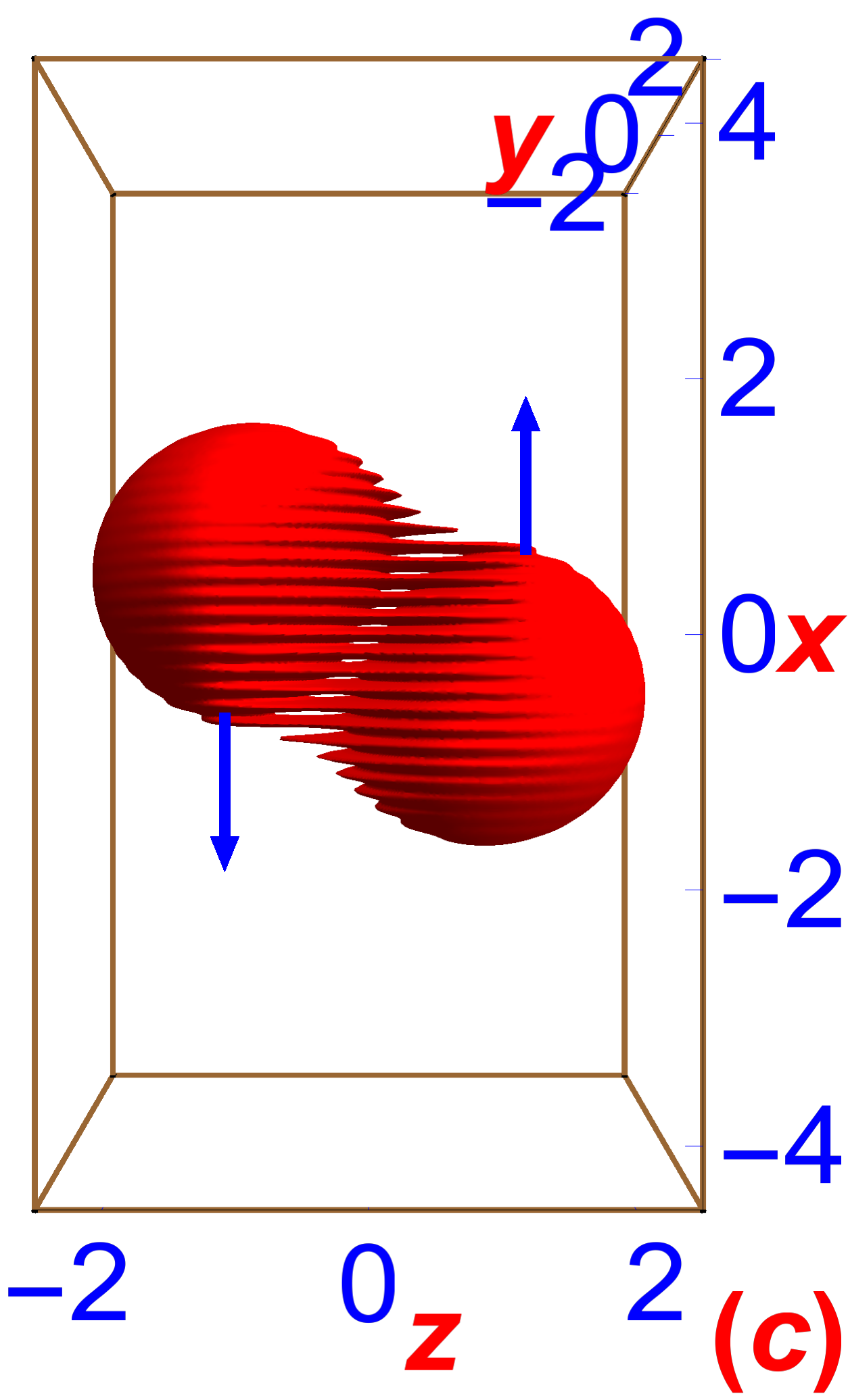}
 \includegraphics[trim = 0mm 0mm 0mm 0mm, clip,width=.32\linewidth]{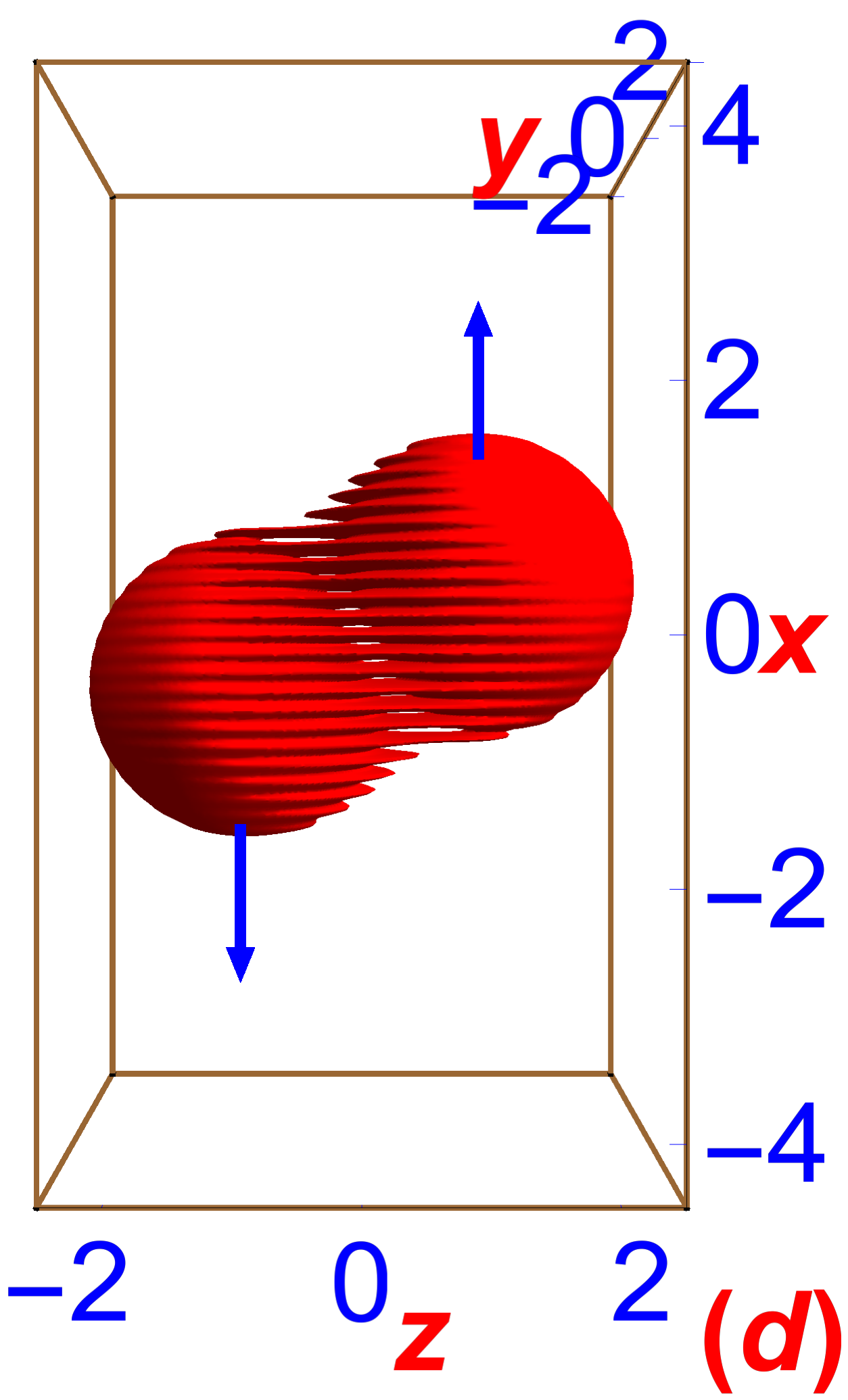}
 \includegraphics[trim = 0mm 0mm 0mm 0mm, clip,width=.32\linewidth]{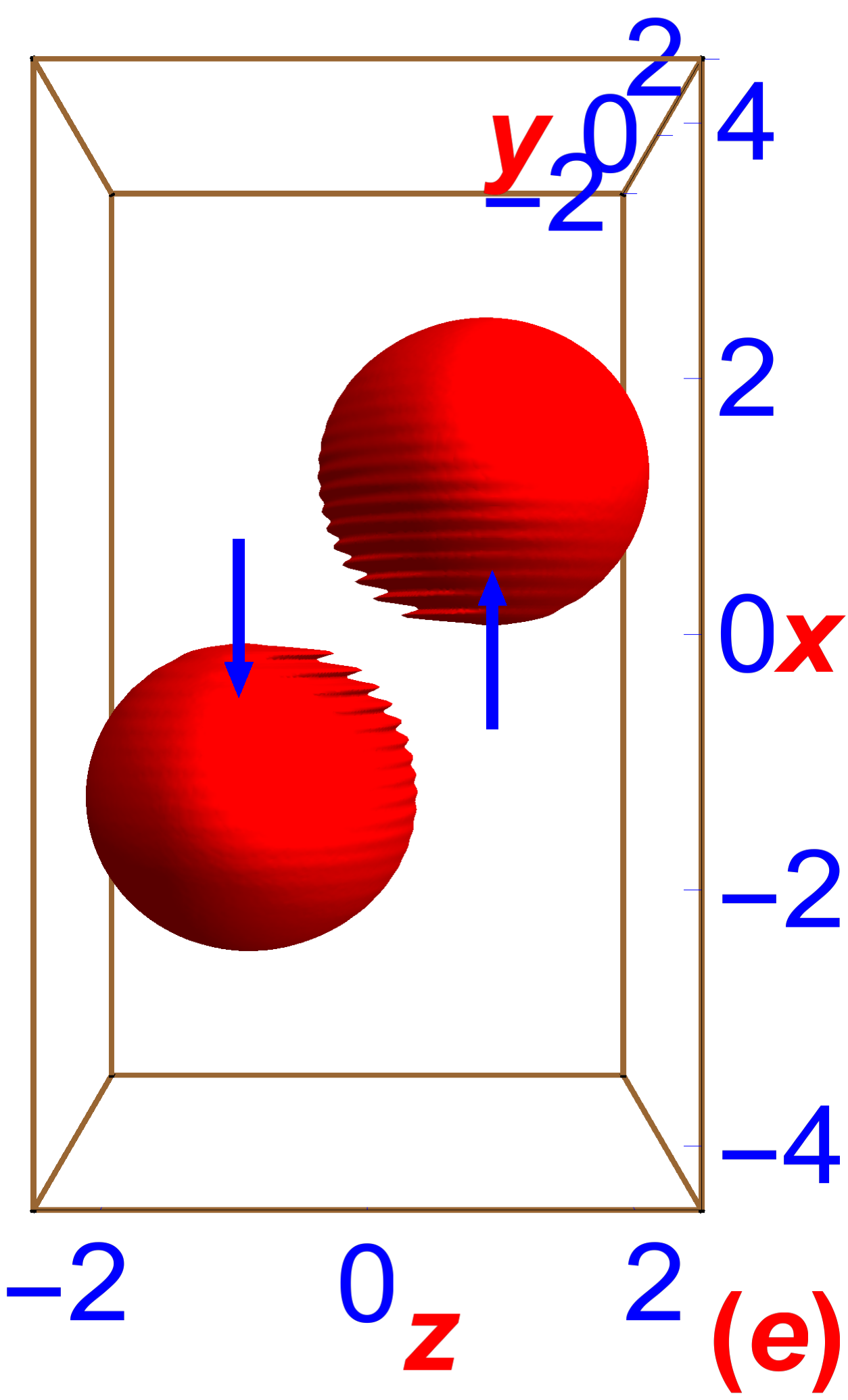} 
\includegraphics[trim = 0mm 0mm 0mm 0mm, clip,width=.32\linewidth]{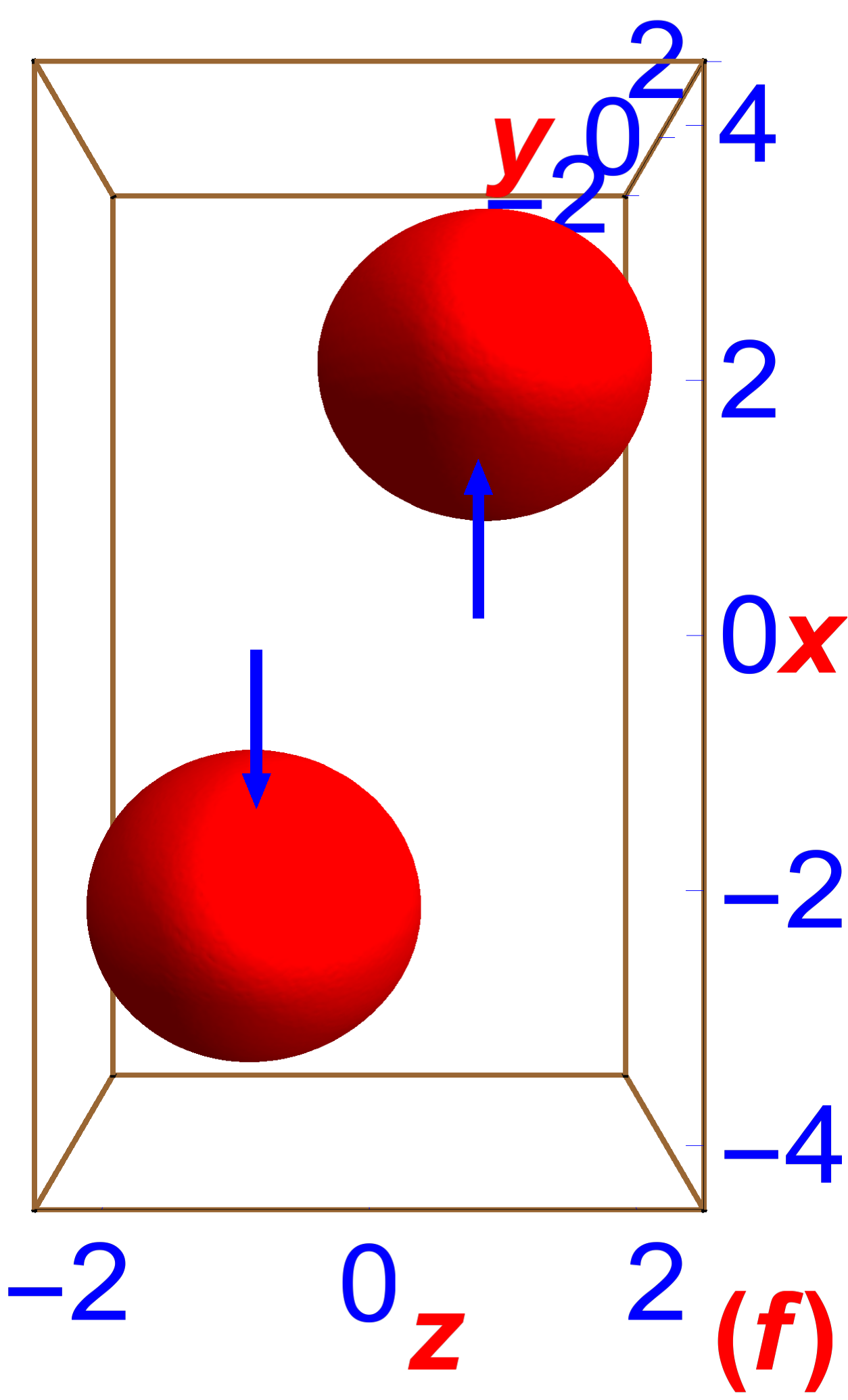}
\caption{ (Color online) Collision dynamics  of two $^7$Li quantum balls, with $N=1500, K_3=3\times 10^{-37}(1-i)$ m$^6$/s each, placed at $x=\pm 2.5$ $\mu$m, $z=\mp 0.9$ $\mu$m at $t=0$ 
 and set into motion in opposite directions along the $x$ axis with velocity  18.16 cm/s, so as to collide at $x=0$, illustrated by   isodensity contours at times  
(a) $t=0$, (b) = 0.0057 ms, (c) = 0.0114 ms, (d) = 0.017 ms, (e) =  0.0228 ms, (f) 
= 0.0285 ms.The density on the contour is 10$^{10}$ atoms/cm$^3$ and unit of length is $\mu$m. The directions of motion  of the  quantum balls are shown by arrows.
  }
\label{fig6} \end{center}

\end{figure}

 \begin{figure}[!t]

\begin{center}
\includegraphics[trim = 0mm 0mm 0mm 0mm, clip,width=.32\linewidth]{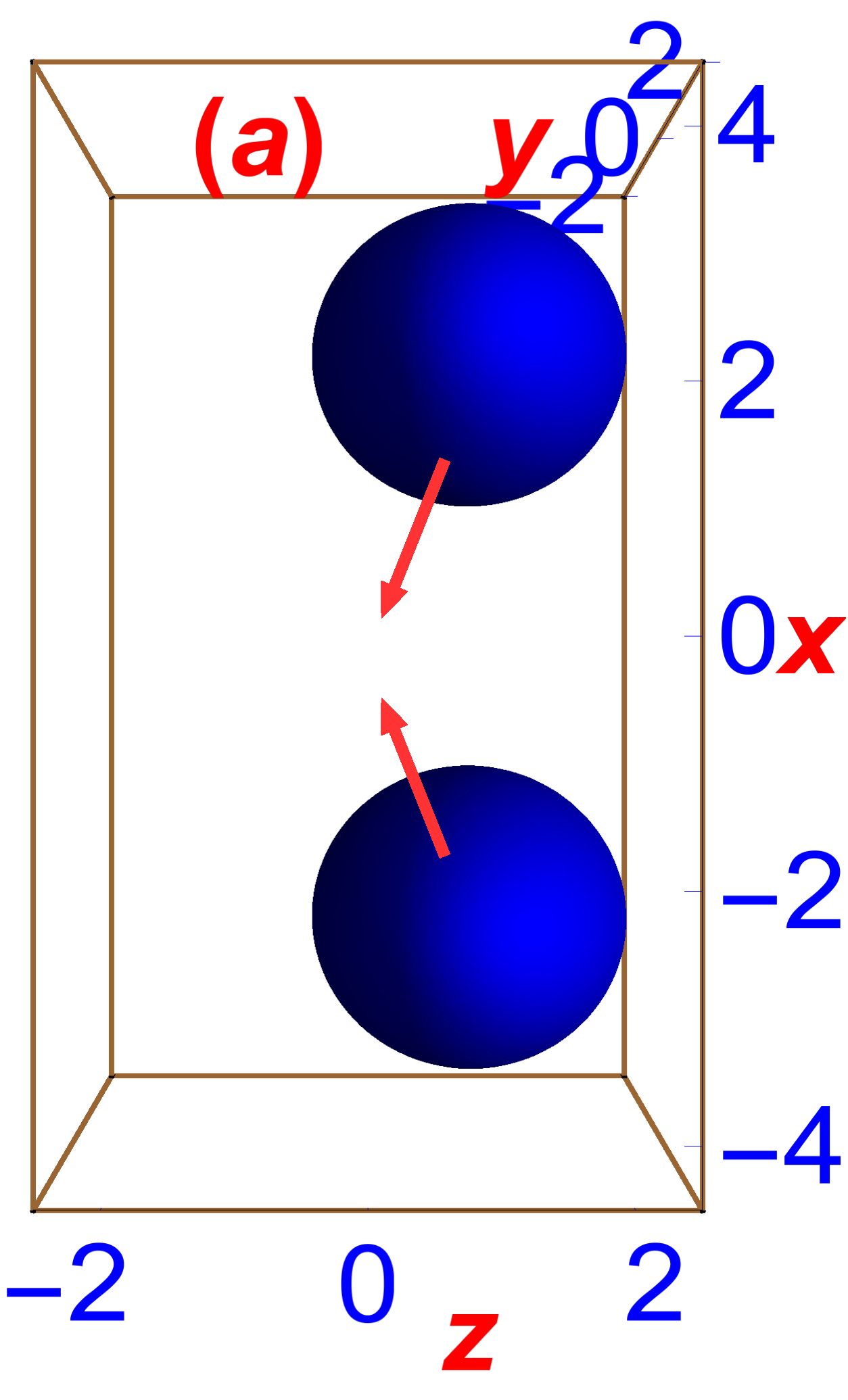}
 \includegraphics[trim = 0mm 0mm 0mm 0mm, clip,width=.32\linewidth]{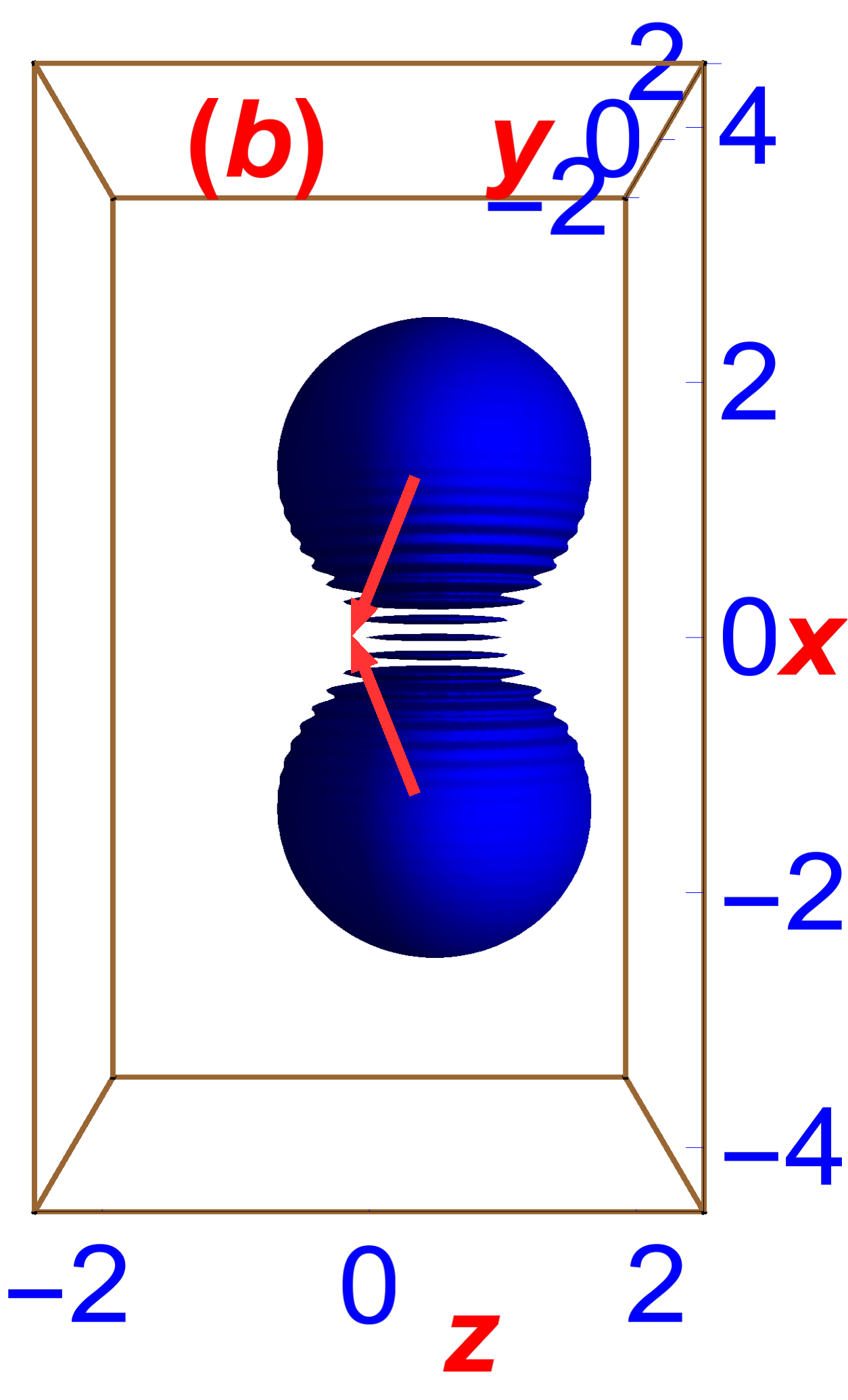} 
\includegraphics[trim = 0mm 0mm 0mm 0mm, clip,width=.32\linewidth]{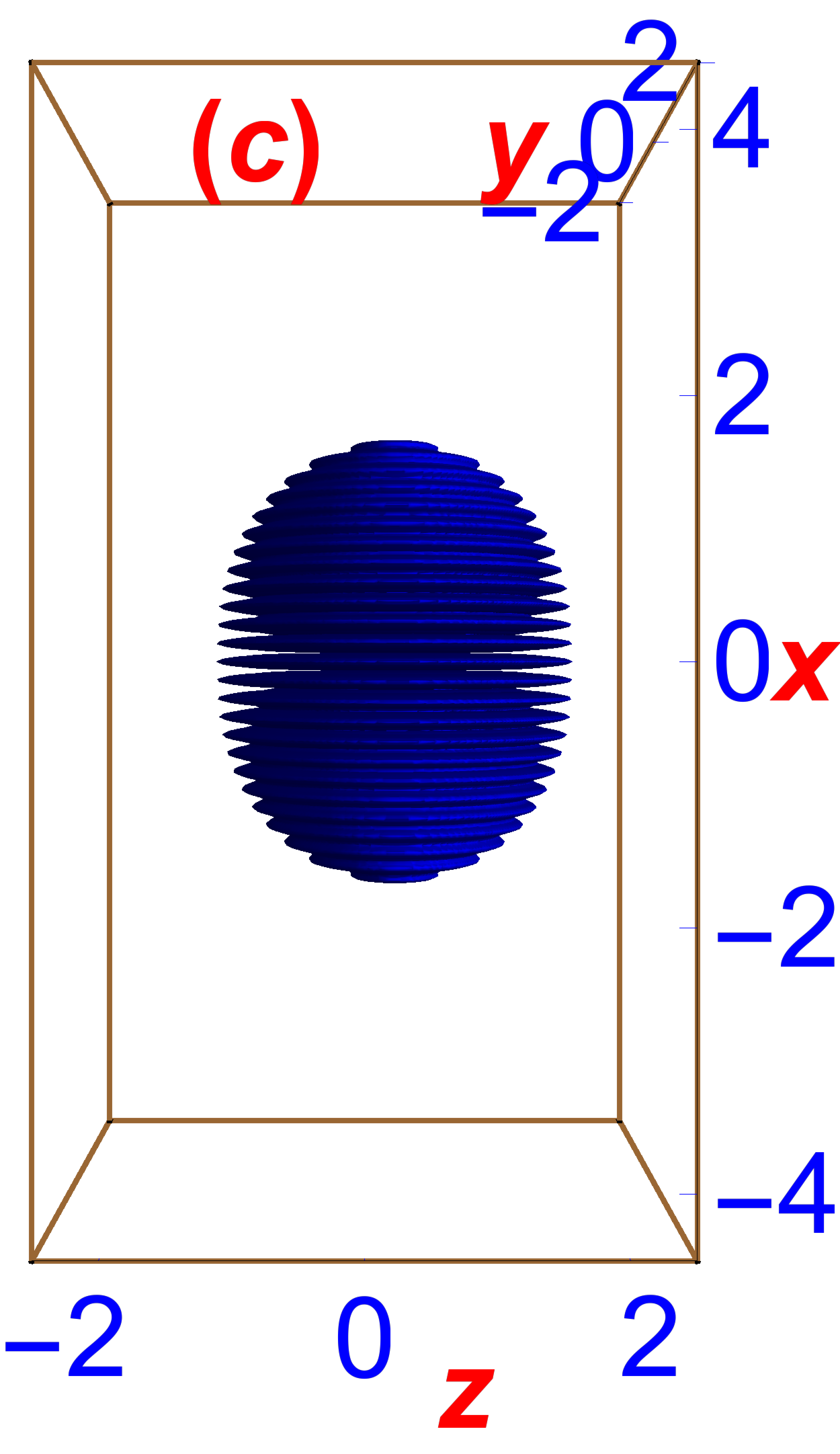}
 \includegraphics[trim = 0mm 0mm 0mm 0mm, clip,width=.32\linewidth]{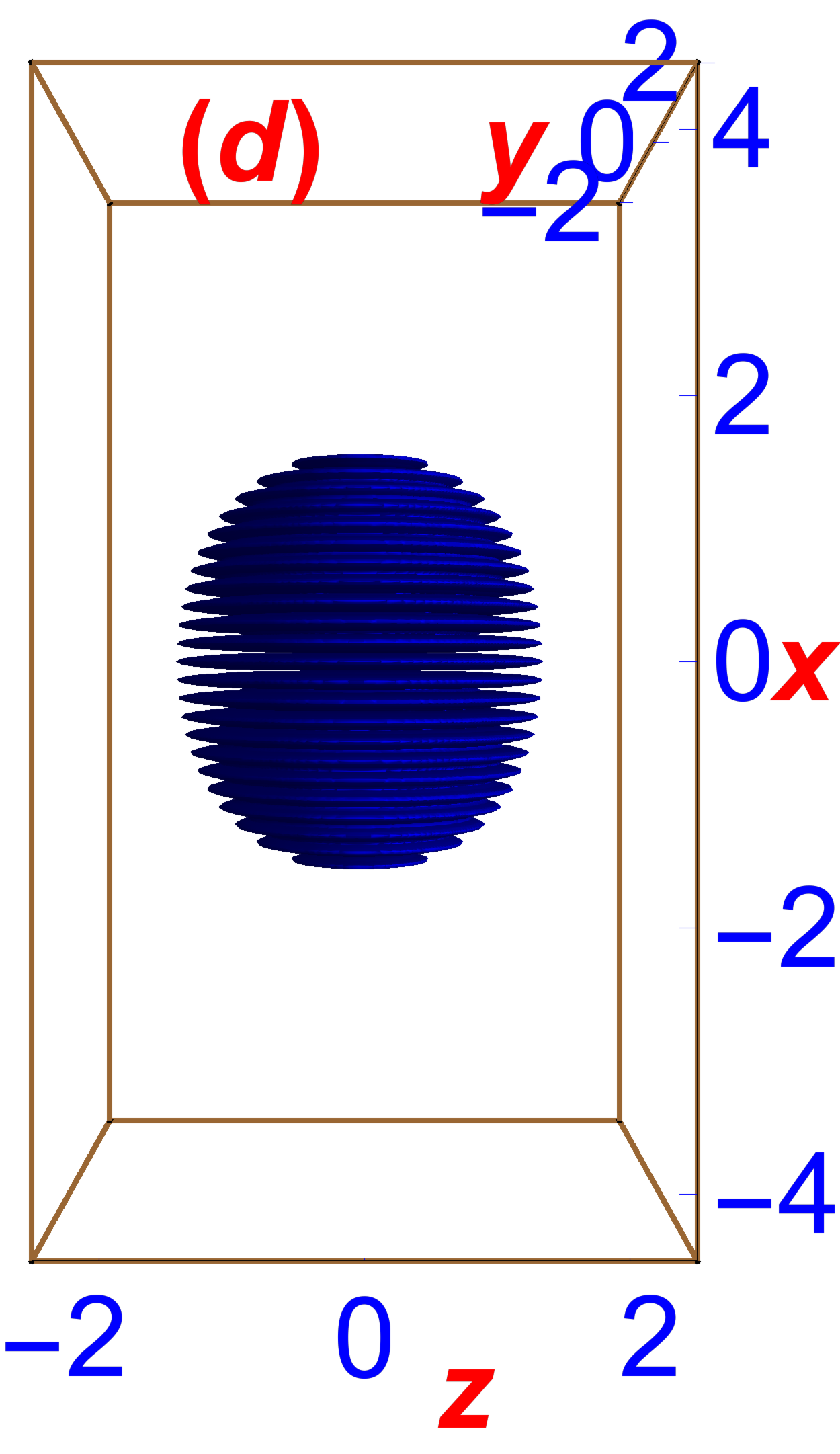}
 \includegraphics[trim = 0mm 0mm 0mm 0mm, clip,width=.32\linewidth]{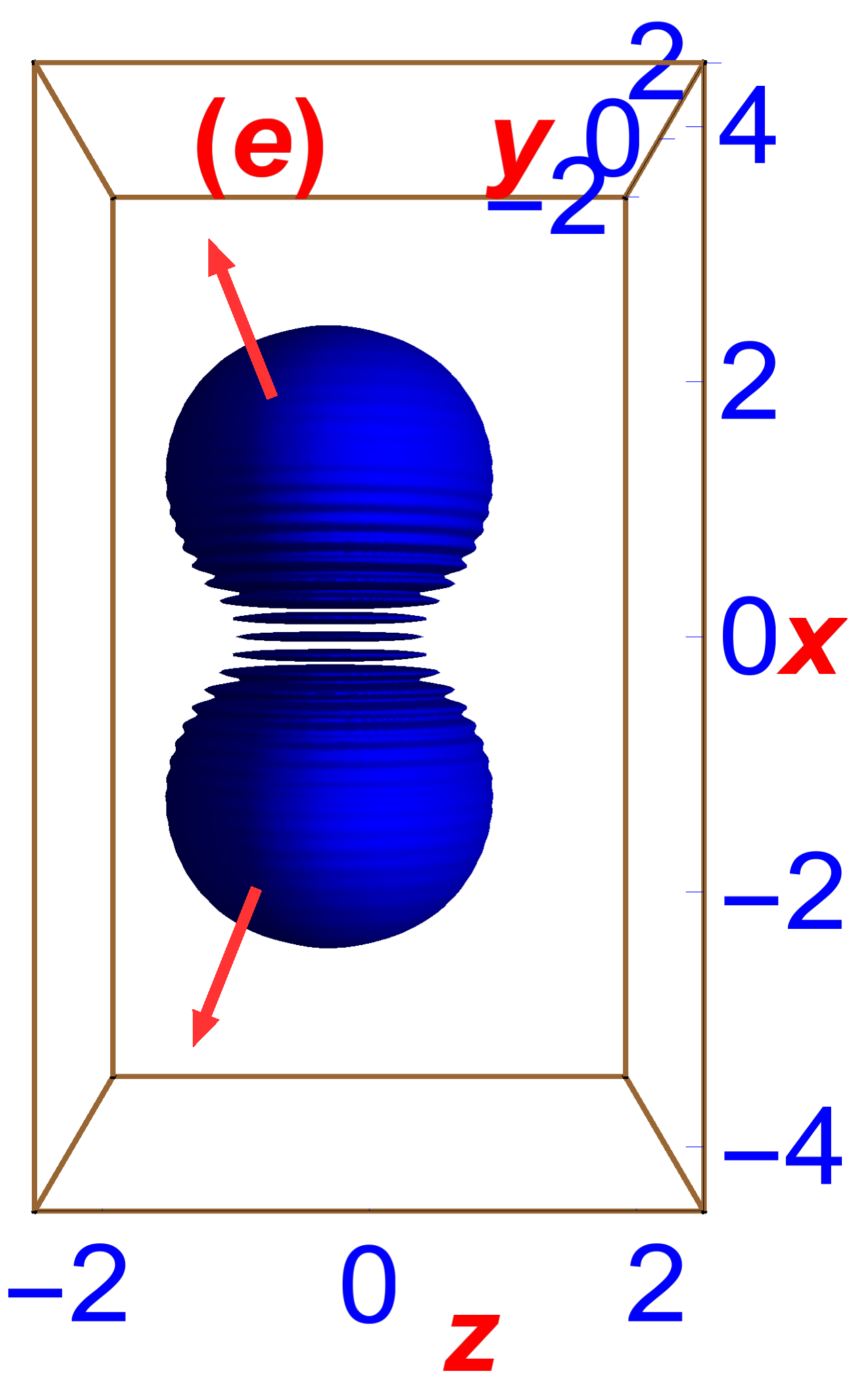} 
\includegraphics[trim = 0mm 0mm 0mm 0mm, clip,width=.32\linewidth]{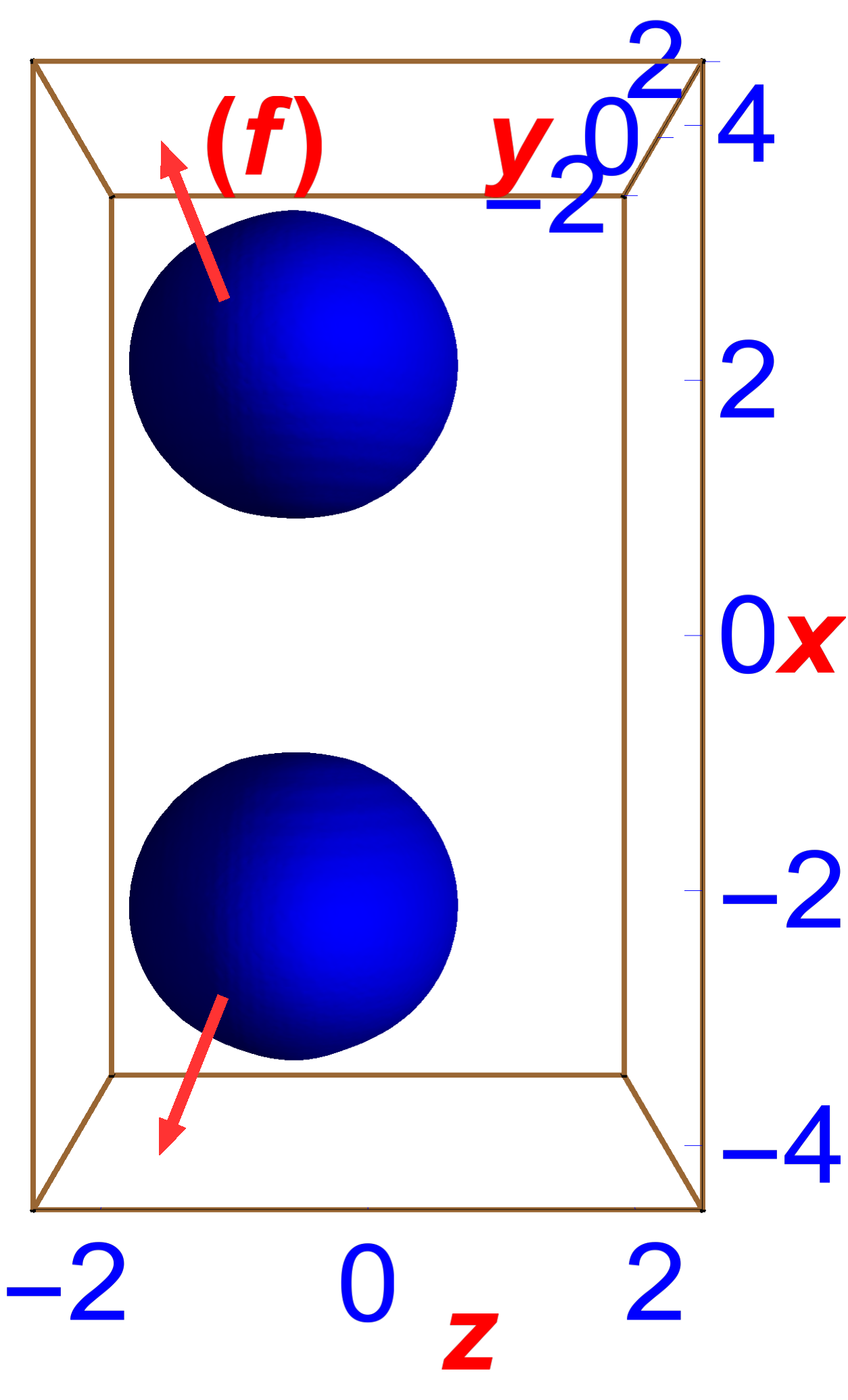}
\caption{ (Color online) Collision dynamics  of two $^7$Li quantum balls, with $N=1500, K_3=3\times 10^{-37}(1-i)$ m$^6$/s each, placed at $x=\pm 2.5$ $\mu$m, $  z= 0.9$ $\mu$m at $t=0$ and set into motion towards origin ($x=z=0$)
  with   velocity  18.9 cm/s,   illustrated by    isodensity contours at times  
(a) $t=0$, (b) = 0.0057 ms, (c) = 0.0114 ms, (d) = 0.017 ms, (e) =  0.0228 ms, (f) 
= 0.0285 ms.  The density on the contour is 10$^{10}$ atoms/cm$^3$ and unit of length is $\mu$m. The directions of motion  of the  quantum balls are shown by arrows.
  }
\label{fig7} \end{center}

\end{figure}

{In the following we study  the  collision  dynamics of  quantum balls, where   we use a three-body term $K_3$ with dissipation corresponding to a loss of atoms from the quantum ball due to molecule formation.  }
To test the solitonic nature of the present  quantum balls,
 we   study the frontal  head-on collision of two quantum balls at large velocity {including a three-body term with absorption: $K_3=3\times 10^{-37}(1-i)$ m$^6$/s.}  
 The imaginary-time profiles of the  quantum balls  shown in  Fig. \ref{fig3} with  $N=1500$ and $K_3=3\times 10^{-37}$ m$^6$/s each  are  used as the initial wave functions in the real-time simulation of collision, with two identical quantum balls  placed at $x=\pm 2.5$  $\mu$m at $t=0$. 
To set the  quantum balls in motion along the $x$ axis in opposite directions the 
respective imaginary-time wave functions are multiplied by $\exp(\pm i 20 x)$ 
  and   real-time simulation is then performed using these wave functions for the study of dynamics. The corresponding    velocity is 
$v=18.16$ cm/s. 
To illustrate the dynamics, we plot  the  isodensity contour of the 
colliding quantum balls in Fig. \ref{fig5} at different times {obtained by real-time simulation over a box of size  $480\times 240 \times 240$.}
The initial profiles of the balls are shown in Fig. \ref{fig5}(a).
 The  balls come close to each other in Fig. \ref{fig5}(b), coalesce to form a single entity in Figs. \ref{fig5}(c) and (d), form two separate balls in Fig. \ref{fig5}(e), 
and are well separated in  Fig. \ref{fig5}(f) with identical profiles as in 
Fig. \ref{fig5}(a). During collision, in Figs.  \ref{fig5}(c) and (d) the identity 
of the two separate balls give rise to a larger object which eventually 
breaks up into two quantum balls. 
 Considering the three-dimensional nature of collision, the distortion in the profile of the quantum balls after collision is found to be negligible, recalling that their 
identities were lost during collision. 
It is useful to contrast the collision shown in Fig. \ref{fig5} with the corresponding 
elastic frontal collision of two classical balls. In the quantum collision the 
identity of the balls is lost during interaction and a quantum ball cannot be followed during collision like a classical ball. Apart from that, the  position and velocity 
of the quantum balls before and after collision are identical with those of the classical balls in this process.  

\begin{figure}[!t]

\begin{center} 
 \includegraphics[trim = 0mm 0mm 0mm 0mm, clip,width=.32\linewidth]{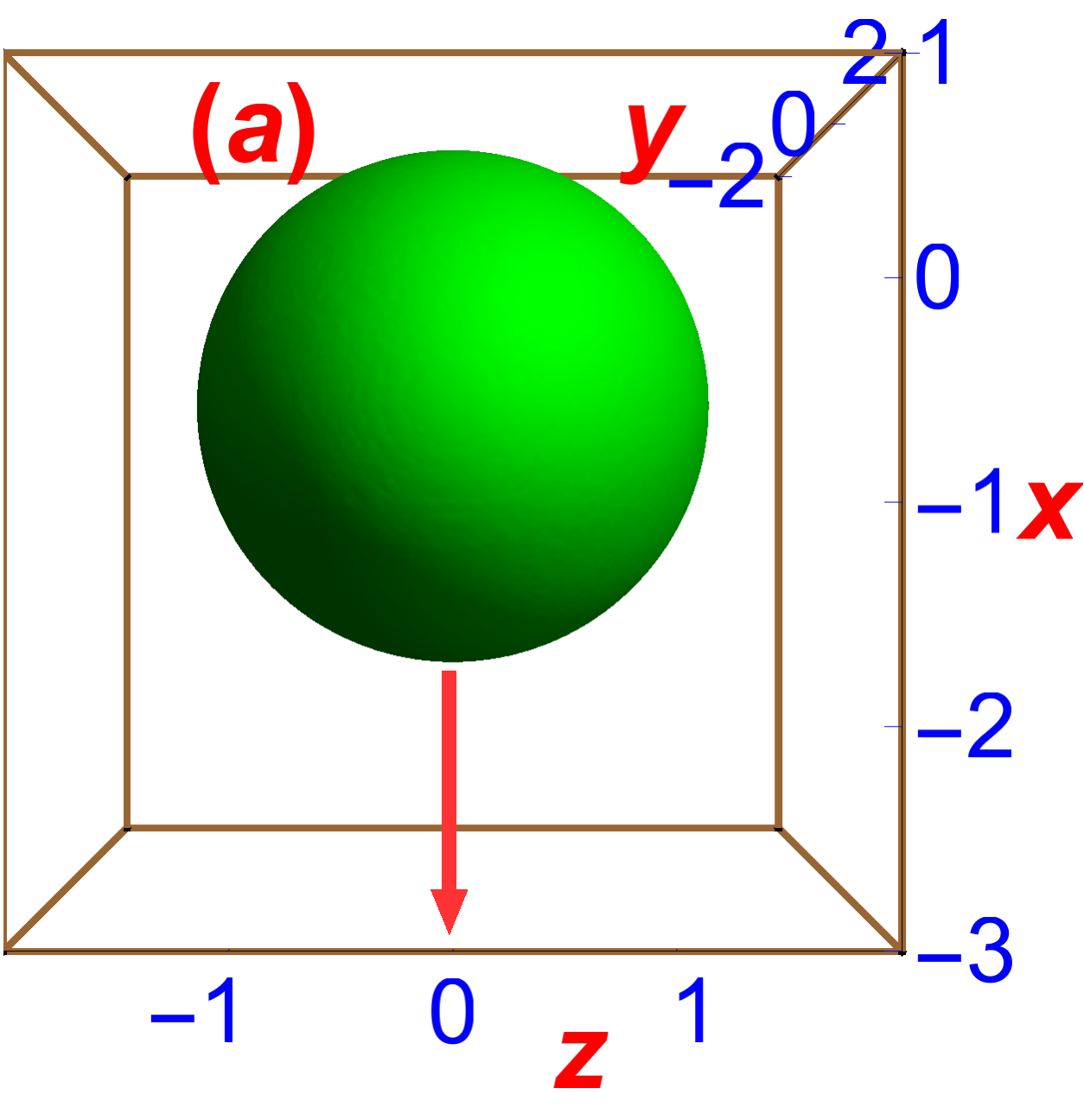}
 \includegraphics[trim = 0mm 2mm 0mm 2mm, clip,width=.32\linewidth]{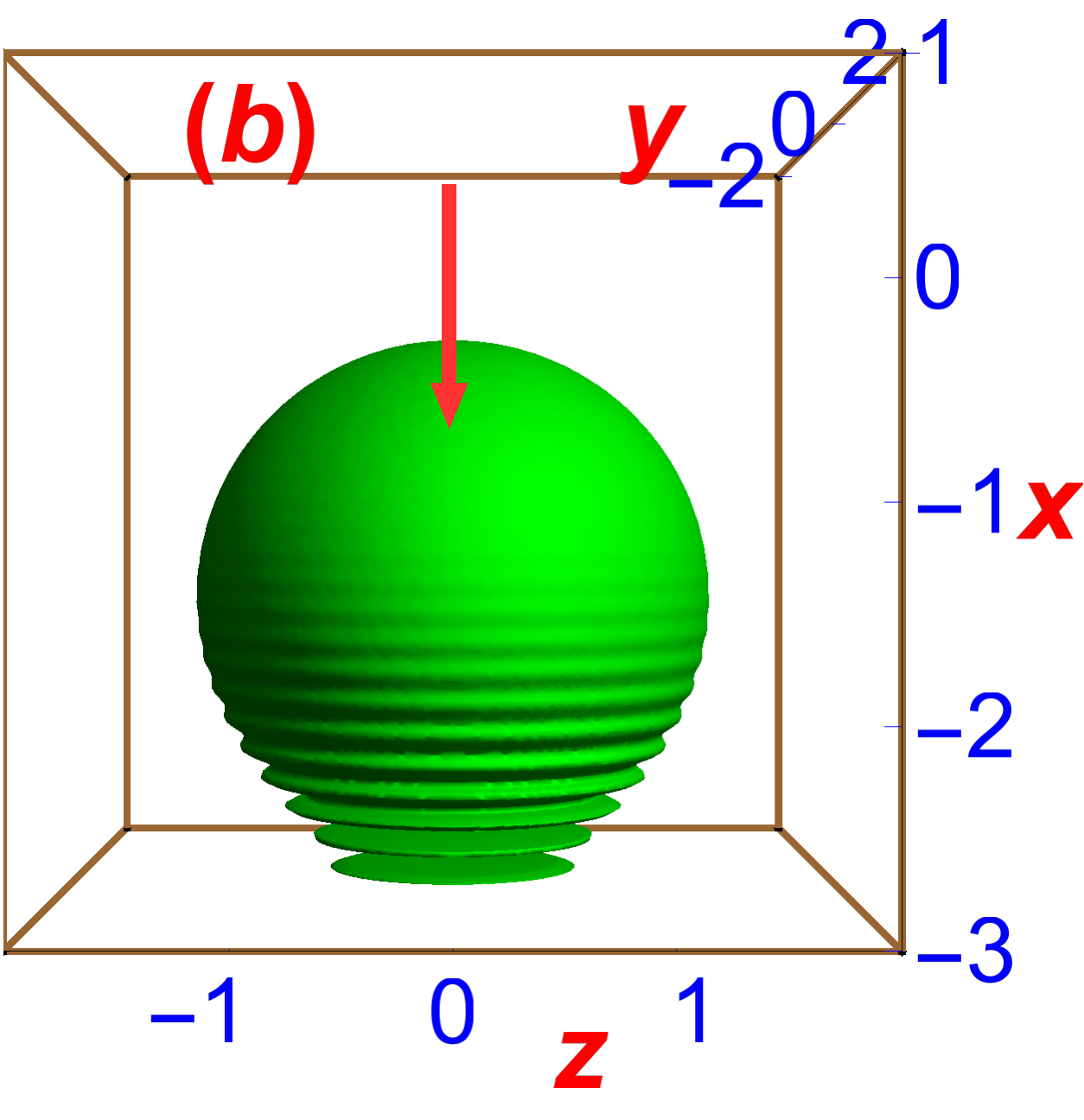} 
\includegraphics[trim = 0mm 0mm 0mm 0mm, clip,width=.32\linewidth]{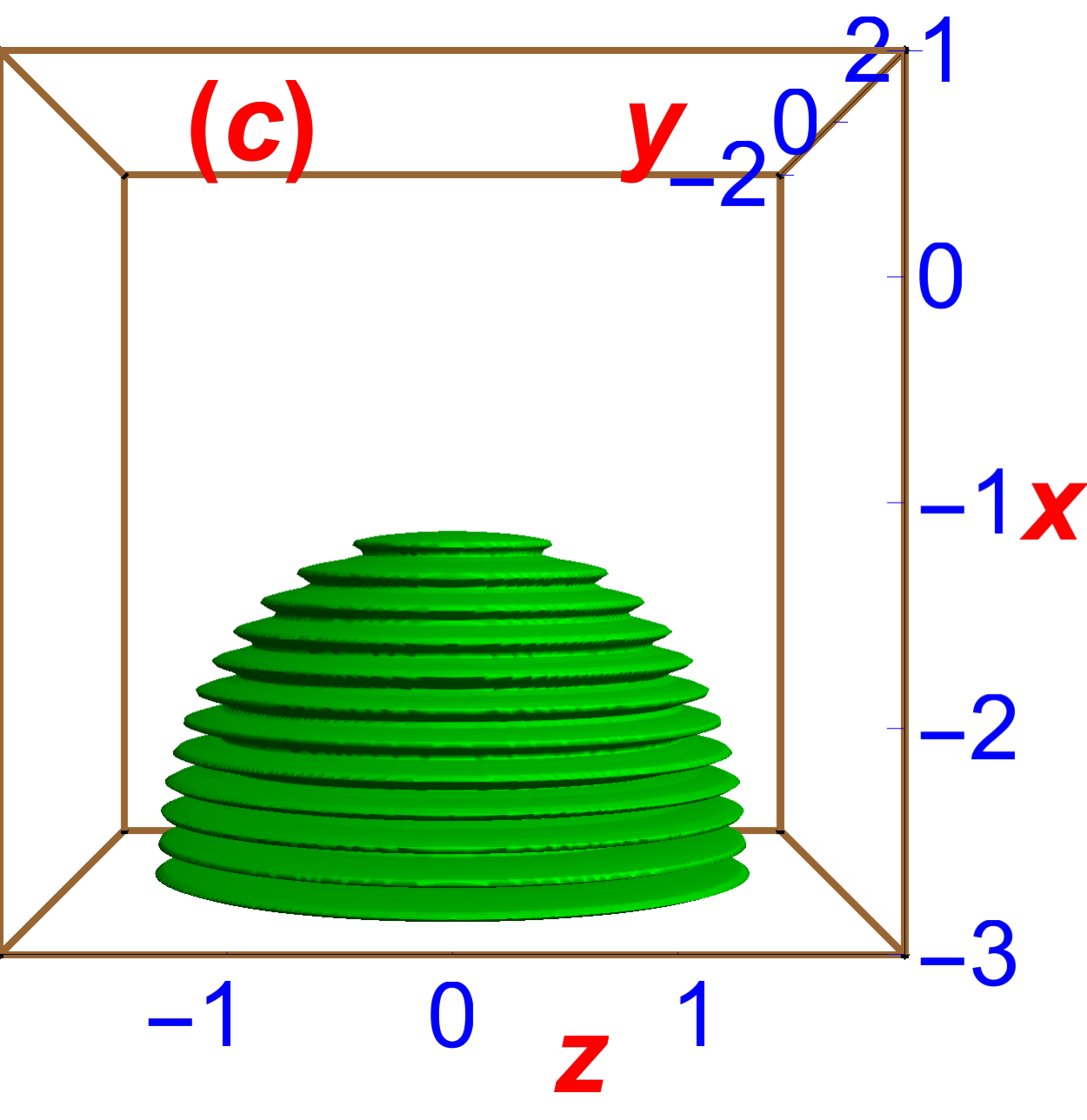}
 \includegraphics[trim = 0mm 0mm 0mm 0mm, clip,width=.32\linewidth]{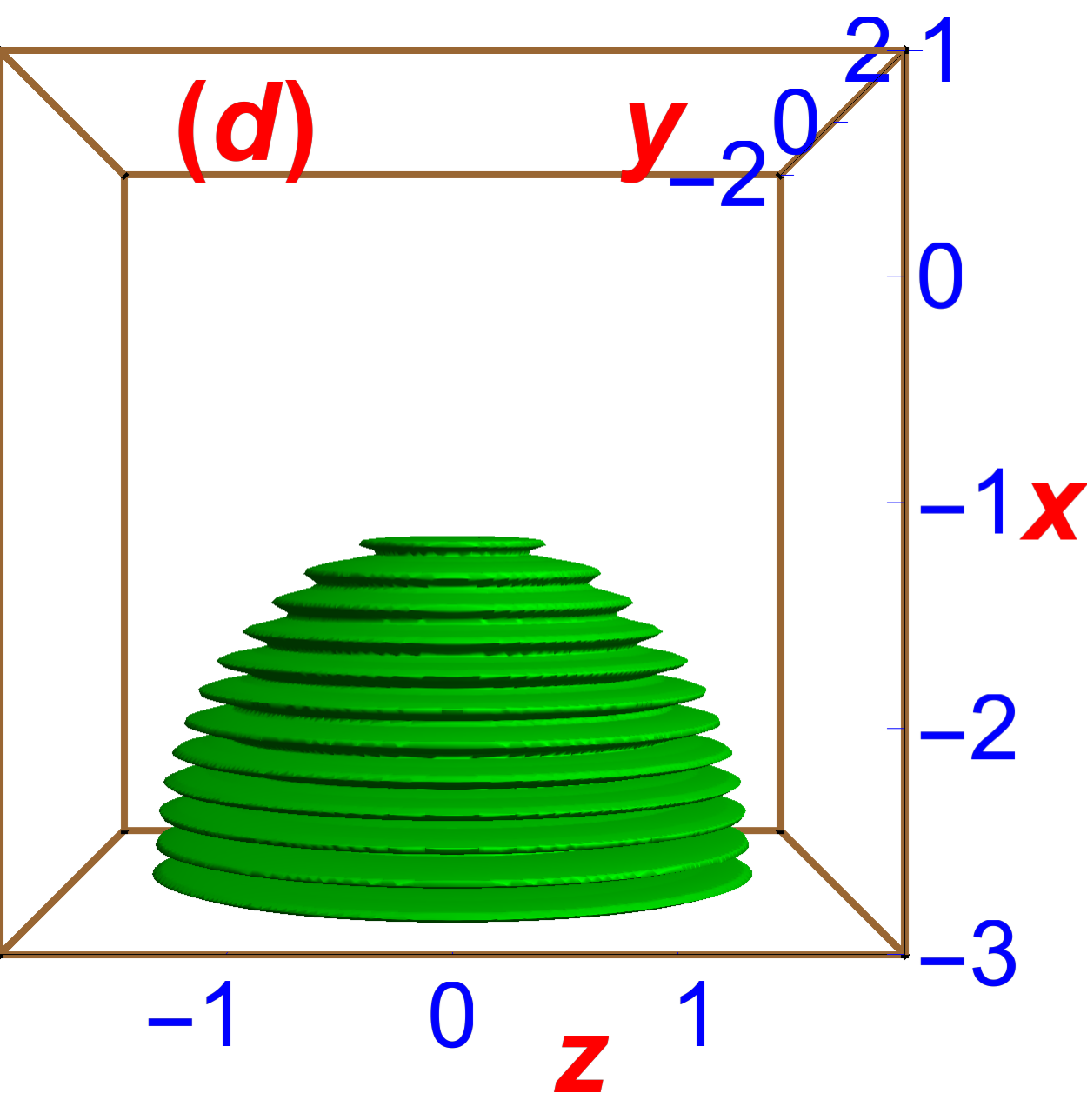}
 \includegraphics[trim = 0mm 2mm 0mm 2mm, clip,width=.32\linewidth]{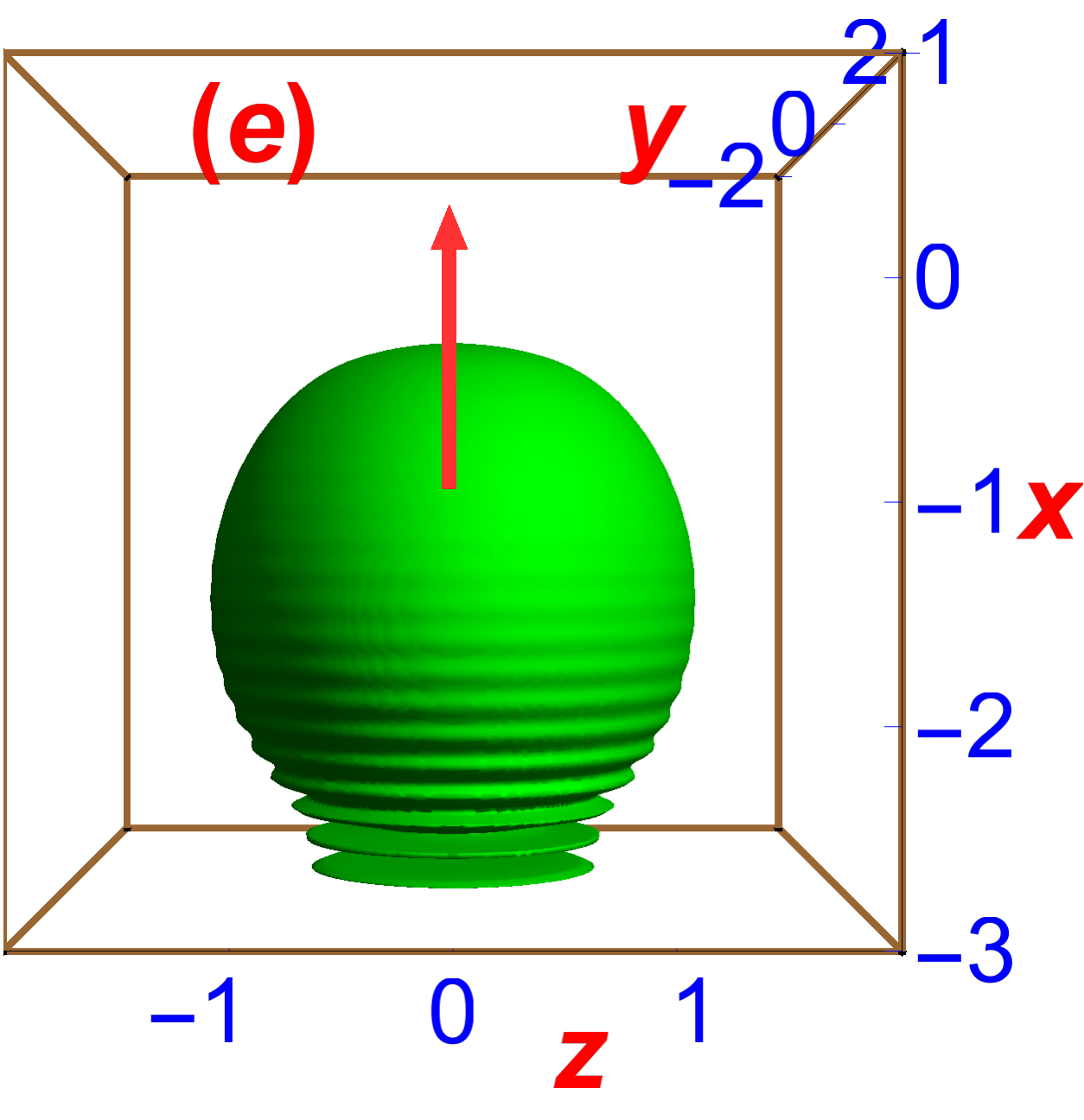} 
\includegraphics[trim = 0mm 0mm 0mm 0mm, clip,width=.32\linewidth]{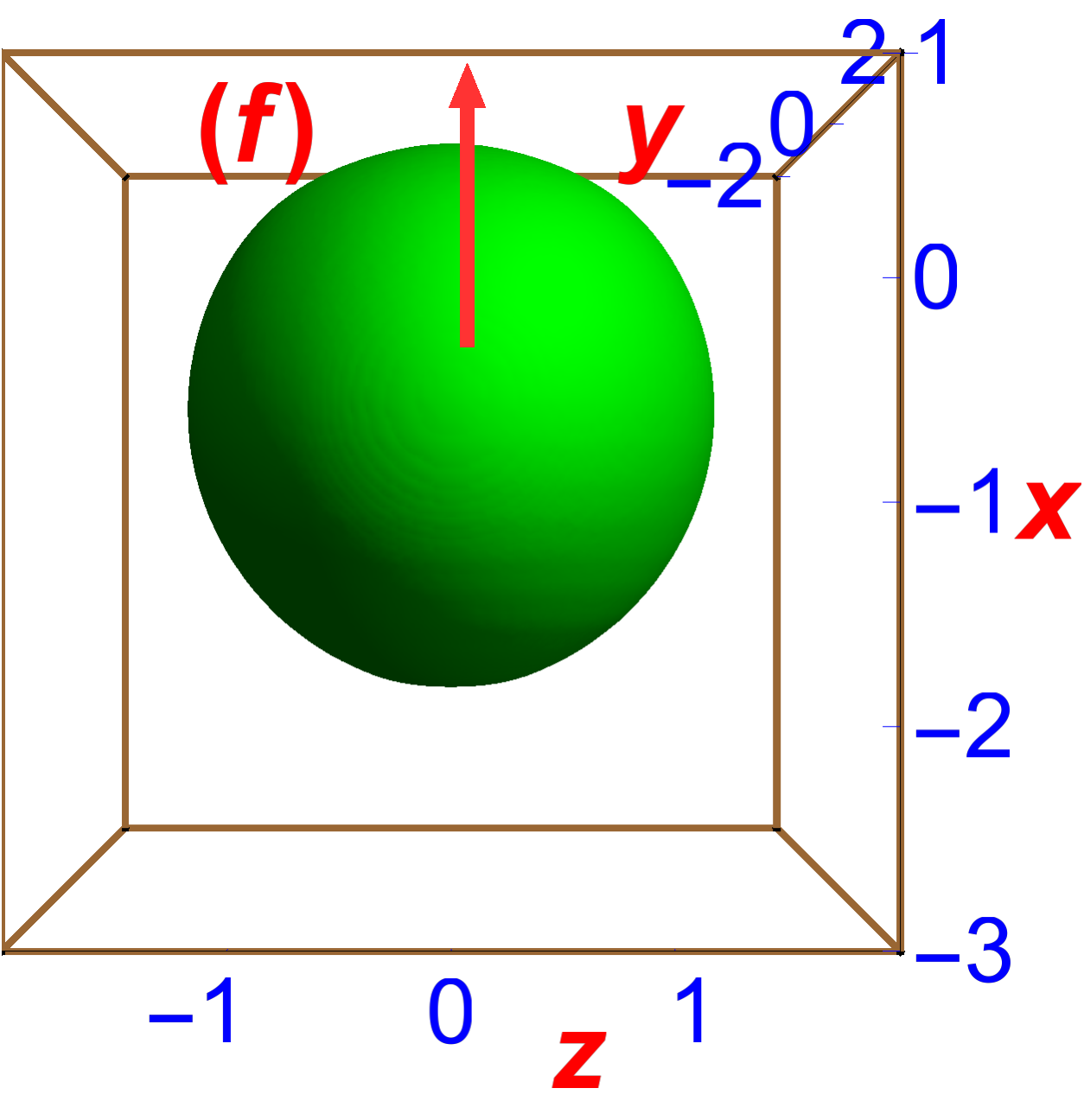}
 
\caption{ (Color online) Bouncing off a rigid wall at $x=-0.5$ $\mu$m of a $^7$Li quantum ball, 
with $N=1500, K_3=3\times 10^{-37}(1-i)$ m$^6$/s,
placed at origin at $t=0$ and moving along $x$ axis with the  velocity of 18.18 cm/s, illustrated 
by 
isodensity contours at times 
(a) $t=0$, (b) = 0.0057 ms, (c) = 0.0114 ms, (d) = 0.017 ms, (e) =  0.0228 ms, (f) 
= 0.0285 ms. The density on
the contour is 10$^{10}$ atoms/cc and unit of length is $\mu$m. The directions of motion of the quantum ball are shown
by an arrows.}
\label{fig8} \end{center}

\end{figure}

Besides the frontal collision considered in Fig. \ref{fig5} we also consider two types  of nonfrontal collisions { including a three-body term with absorption}. First, we consider the collision between two quantum balls moving along the $x$ axis but on laterally displaced tracks. At $t=0$ two balls of  $N=1500$ and $K_3=3\times 10^{-37}(1-i)$ m$^6$/s  each 
are placed at $x=\pm 2.5$ $\mu$m, $y=0, z=\mp 0.9$ $\mu$m, respectively, and set into motion along $x$ axis in opposite directions by multiplying the respective imaginary-time 
wave 
functions by  $\exp(\pm i20 x)$ and performing real-time simulation {in a box of size $480\times 240 \times 360 $} with these wave functions for the study of dynamics. The collision is illustrated in Fig. \ref{fig6}(a)-(f) through successive snapshots of isodensity contours 
of the system before, during and after collision. In this case in Figs. \ref{fig6}(a) and (b)
the balls approach along the $x$ axis, in Figs. \ref{fig6}(c) and (d) they
interact by losing identity, and in Figs. \ref{fig6}(e) and (f) they eventually 
 come out of the interaction region undeformed while moving along the $x$ axis
maintaining their original trajectories and  conserving their velocities.  This collision has no classical analogue. In the elastic collision  of two classical balls in this case, the balls will be deflected from their original trajectories conserving energy and momentum.

 Another type of collision of interest is the angular collision of two quantum balls which we now study.  For this purpose, at $t=0$  two balls of  $N=1500$ and $K_3=3\times 10^{-37}(1-i)$ m$^6$/s  each 
are placed at $x=\pm 2.5$ $\mu$m, $y=0, z=0.9$ $\mu$m, respectively, and set into motion towards the origin $x=y=z=0$ with equal velocities by multiplying the respective imaginary-time 
wave functions 
by  $\exp(\pm i 20 x+i 5.8 z)$  to set the quantum balls in motion with a 
velocity of 18.9 cm/s and performing real-time simulation {in a box of size  $ 480\times 240\times 360$} with these wave functions for the study of dynamics. 
Again the isodensity profiles of the quantum balls before, during, and after collision are shown in Figs. \ref{fig7}(a)-(b), (c)-(d), and 
(e)-(f), respectively. In this case the balls again come out after collision undeformed maintaining their original trajectories and  conserving their velocities.
 If we contrast this collision with the corresponding  elastic collision of two classical balls,  the  position and velocity 
of the quantum balls before and after collision are identical with those of the classical balls in this process.  However, again the quantum balls lose their identity during the collision.

The elastic interaction of a quantum ball with external objects is also of interest.
For this purpose we consider its interaction with a rigid elastic plane upon perpendicular and angular impacts { including a three-body term with absorption}.  To study the vertical impact with a rigid elastic plane 
we place a quantum  ball of  $N=1500$ and $K_3=3\times 10^{-37}(1-i)$ m$^6$/s 
at   $x=-0.5, $ and Q$y=z=0$  at $t=0$ and set it in motion along the $x$ axis with a velocity of 18.18 cm/s by multiplying 
the imaginary-time  
wave function by $\exp(-i20 x)$ 
and performing real-time simulation with this wave function {in a box of size $240\times 240\times 240$. The reflection of the ball from the wall 
is  achieved by just imposing reflecting boundary condition at the surface in the Crank-Nicolson algorithm \cite{CPC}.}
   The interaction dynamics in this case is illustrated by successive snapshots of isodensity contour before, during and after 
interaction with the rigid plane in Figs. \ref{fig8}(a)-(f). The quantum ball moves without any deformation in  Figs. \ref{fig8}(a)-(b), gets deformed in proximation of the rigid plane in Figs. \ref{fig8}(c)-(d), bounces off without any deformation and without any 
change of speed  and with the direction of motion reversed
in Figs. \ref{fig8}(e)-(f).  The dynamics of the quantum ball is the same as that of an elastic classical ball  except near the rigid plane when the quantum ball gets deformed.

 \begin{figure}[!t]

\begin{center} 
 \includegraphics[trim = 0mm 0mm 0mm 0mm, clip,width=.32\linewidth]{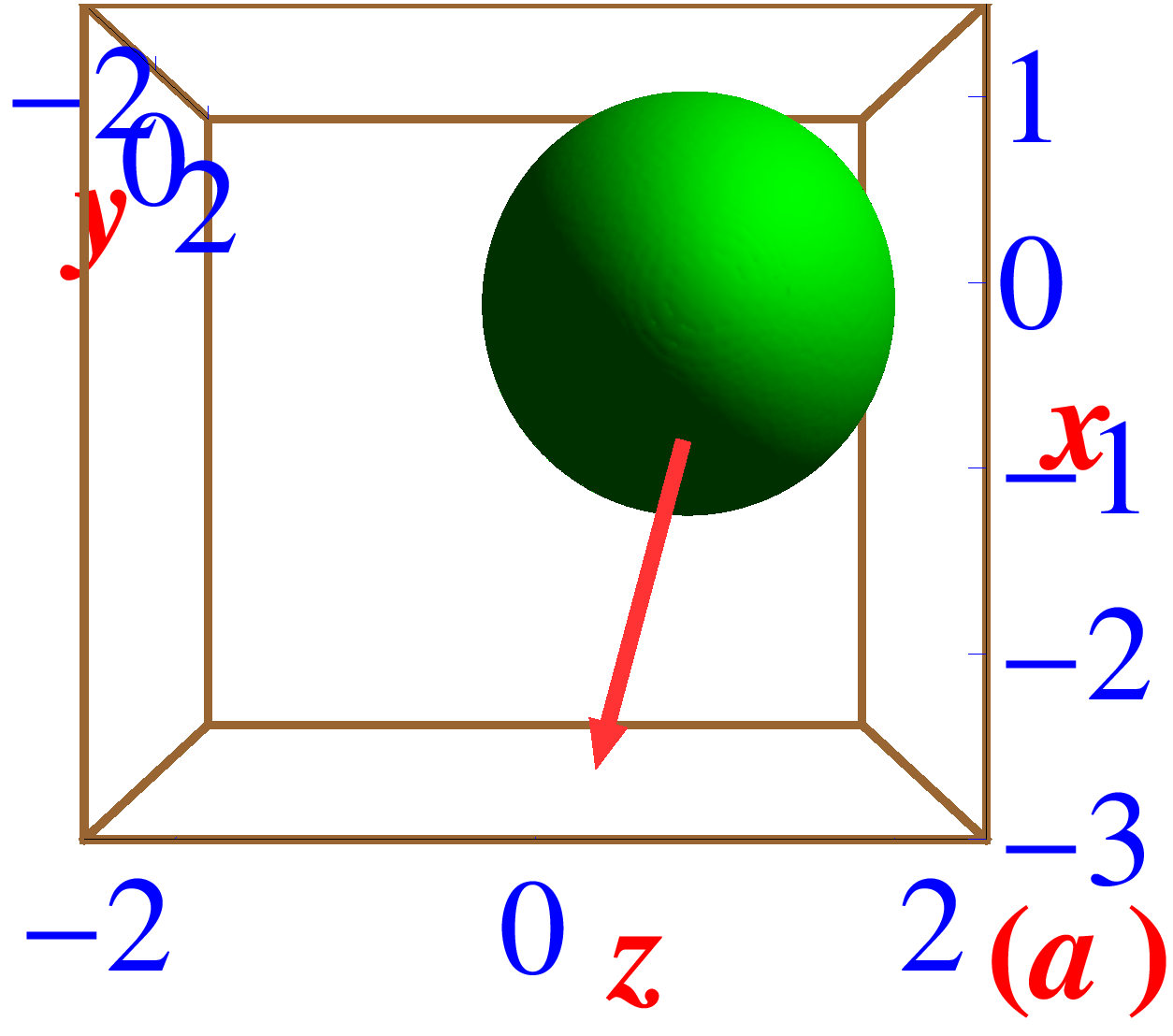}
 \includegraphics[trim = 0mm 0mm 0mm 0mm, clip,width=.32\linewidth]{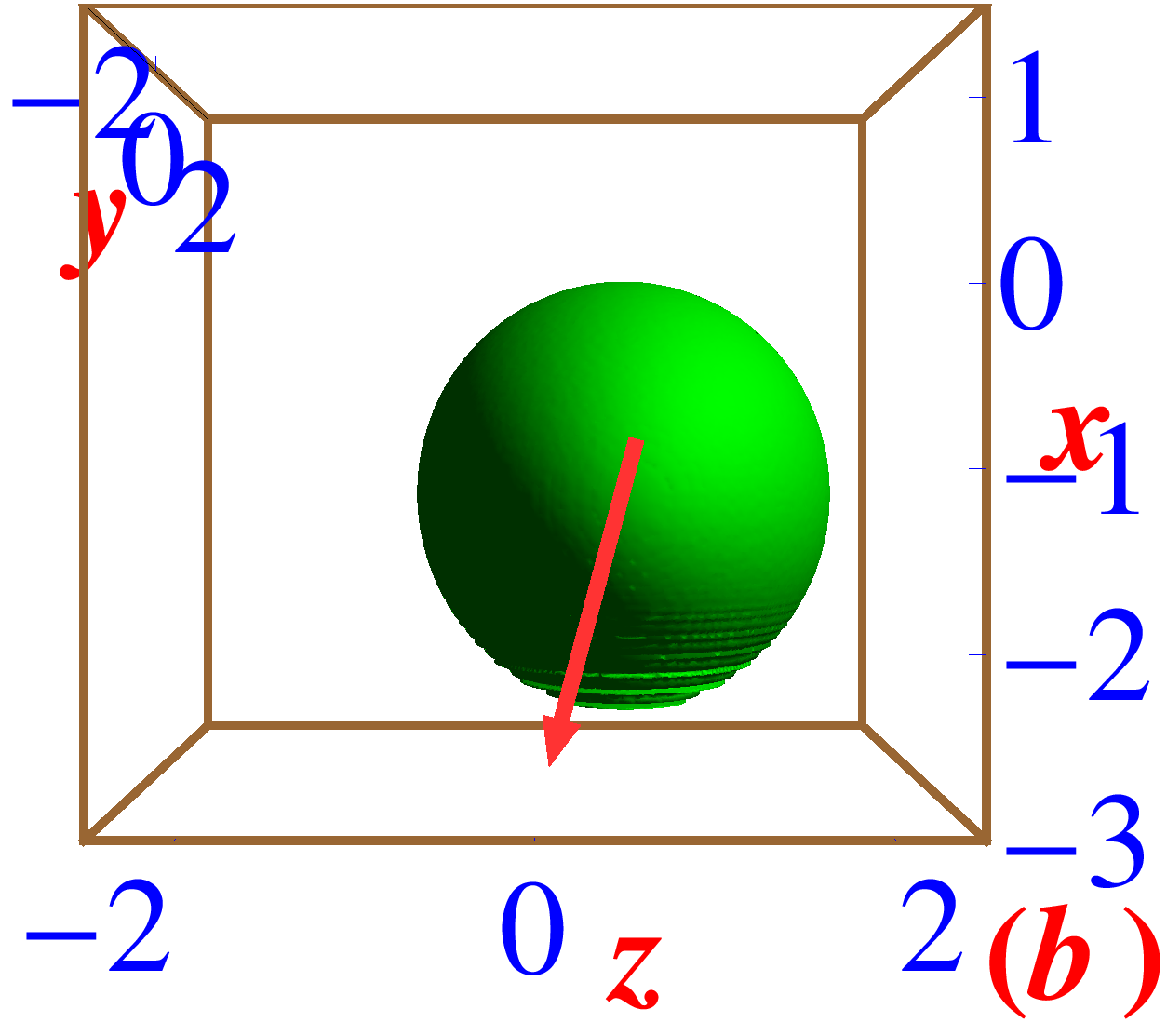} 
\includegraphics[trim = 0mm 0mm 0mm 0mm, clip,width=.32\linewidth]{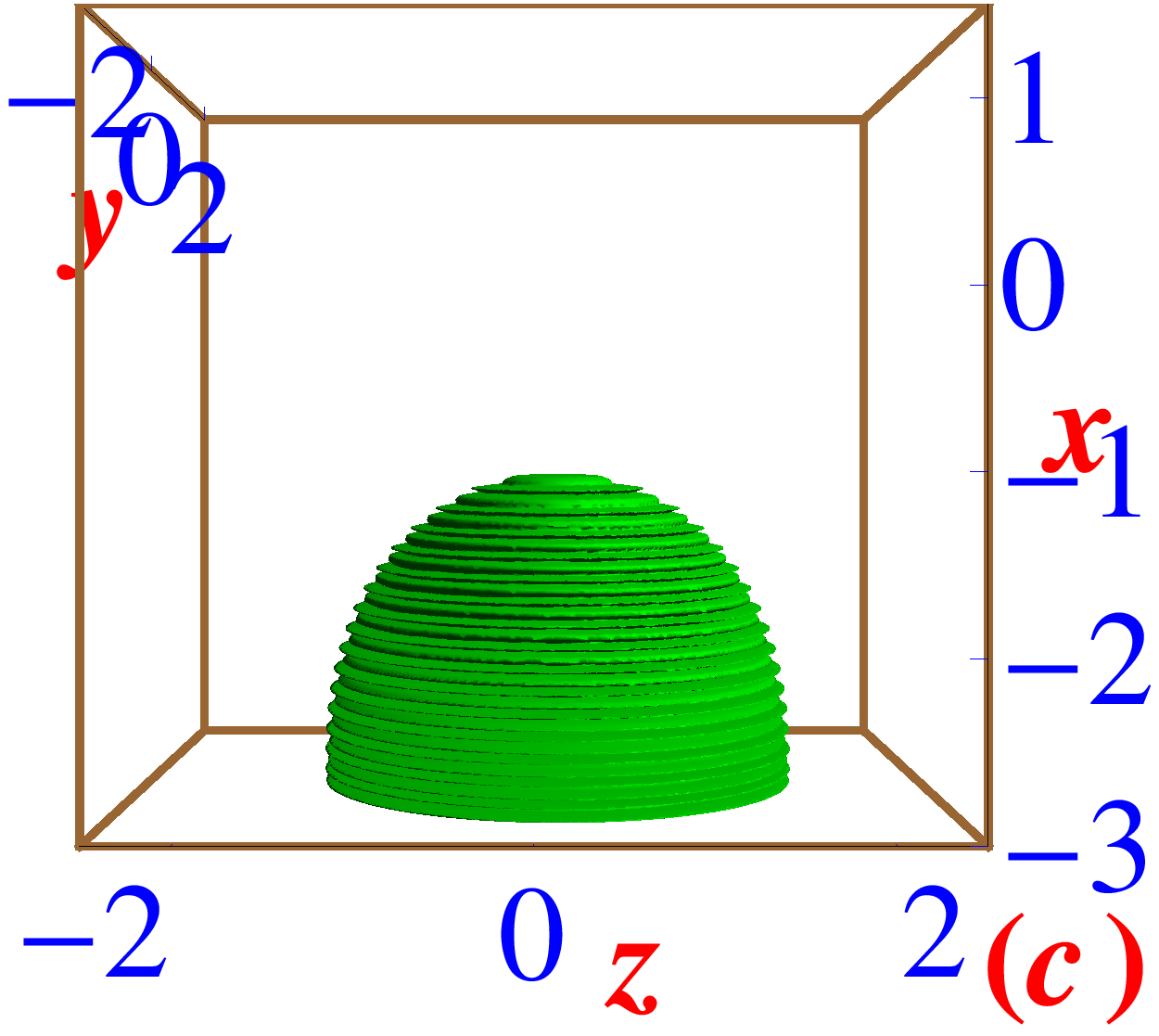}
 \includegraphics[trim = 0mm 0mm 0mm 0mm, clip,width=.32\linewidth]{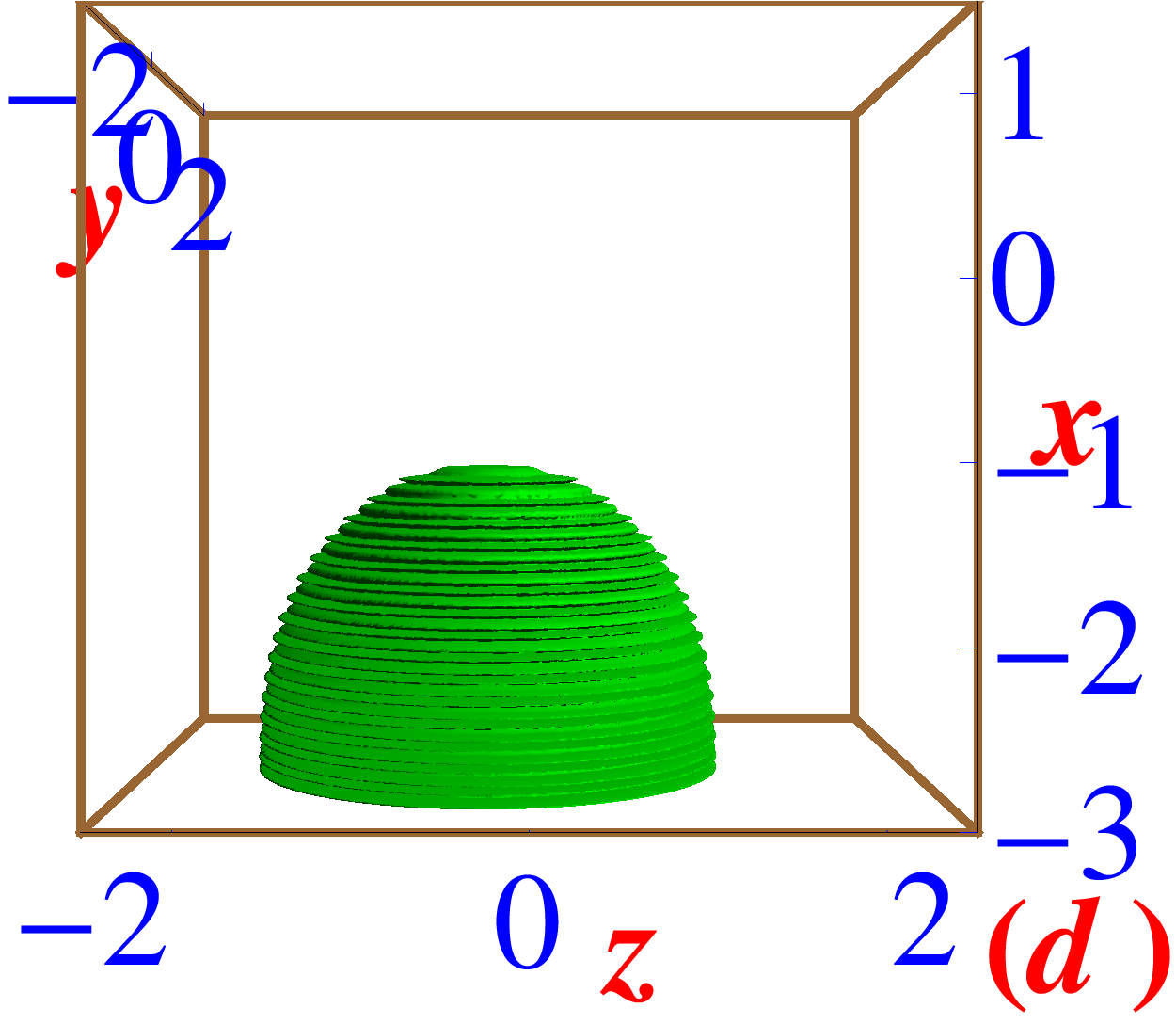}
 \includegraphics[trim = 0mm 0mm 0mm 0mm, clip,width=.32\linewidth]{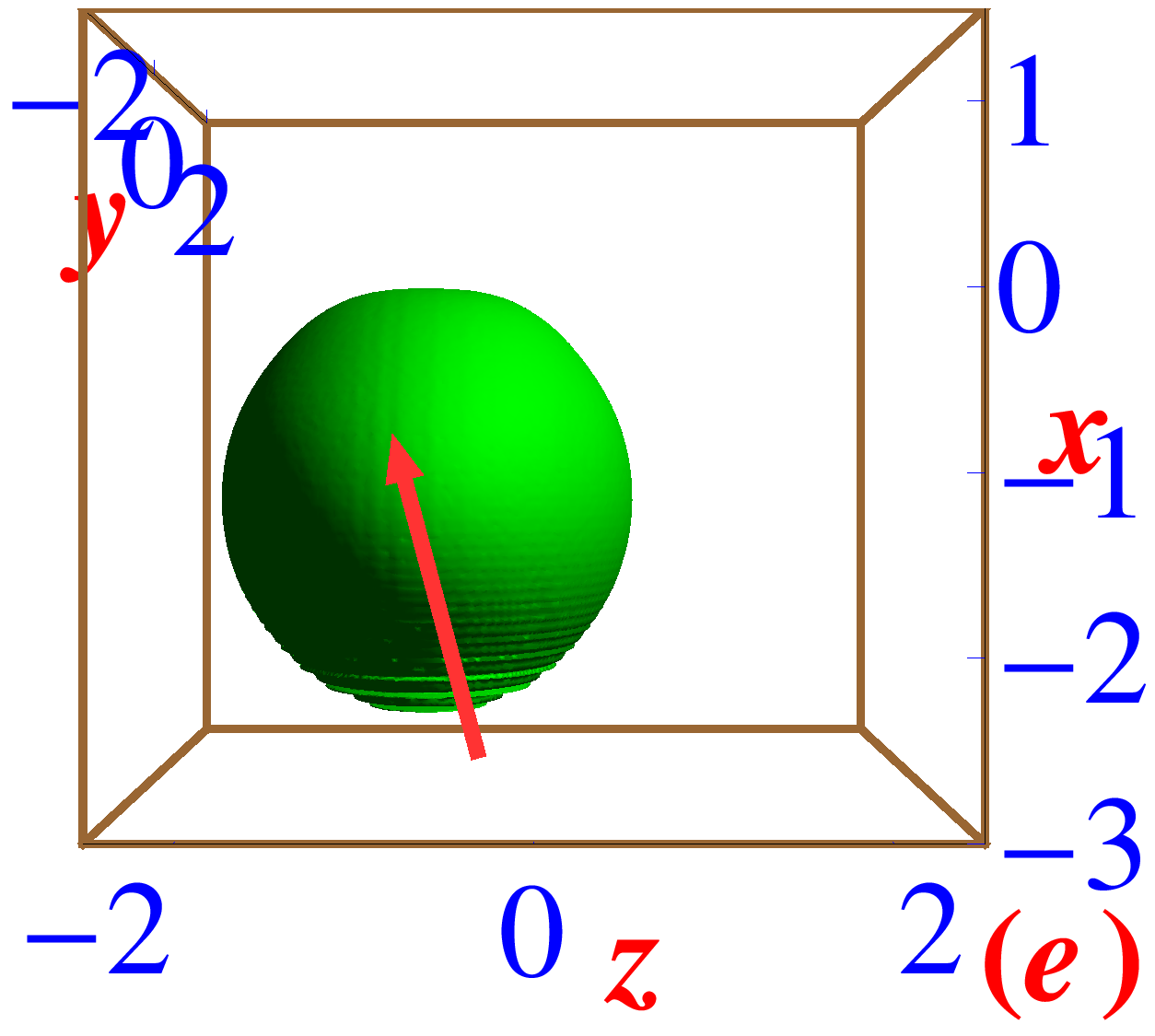} 
\includegraphics[trim = 0mm 0mm 0mm 0mm, clip,width=.32\linewidth]{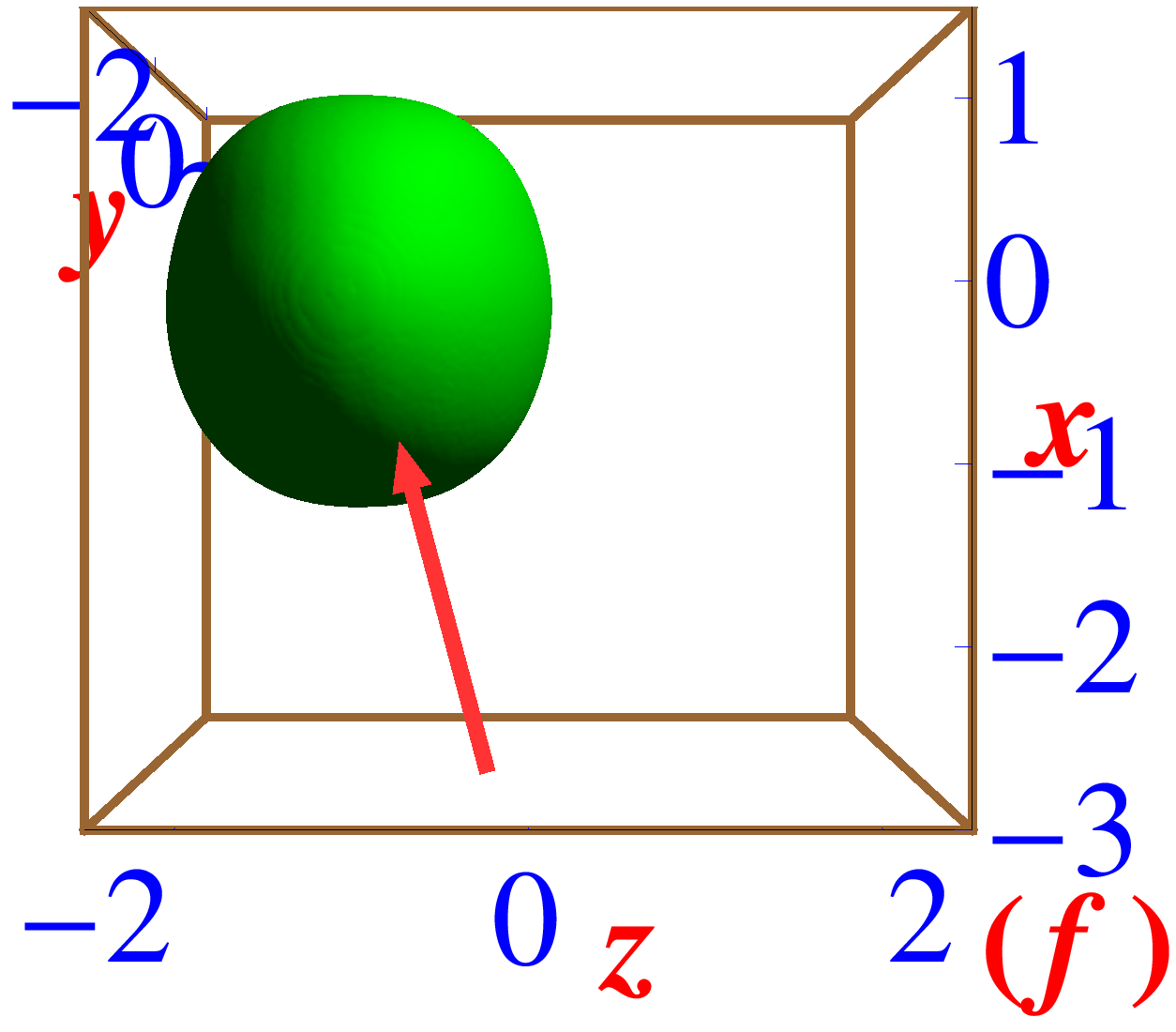} 
\caption{ (Color online)
Bouncing off a rigid wall at $x=-3$ $\mu$m of a $^7$Li quantum ball, 
with $N=1500, K_3=3\times 10^{-37}(1-i)$ m$^6$/s,
placed at $x=0,z=1$ $\mu$m at $t=0$ and set into motion towards $x=-3$ $\mu$m,  $z=0$
 with   velocity 18.9 cm/s,   illustrated 
by dimensionless
isodensity contours at times 
(a) $t=0$, (b) = 0.00669 ms, (c) = 0.0134 ms, (d) = 0.020 ms, (e) =  0.0267 ms, (f) 
= 0.0334 ms.  The density on the contour is 10$^{10}$ atoms/cm$^3$. The directions of motion of the quantum ball are shown
by  arrows. }
\label{fig9} \end{center}

\end{figure}

The interaction of a quantum ball upon angular impact with a rigid plane is studied 
next.  A quantum ball of  $N=1500$ and $K_3=3\times 10^{-37}(1-i)$ m$^6$/s is placed at 
$x=y=0, z=1$ $\mu$m and set into motion towards $x=-3$ $\mu$m, $y=z=0$ by multiplying the imaginary-time wave function by $\exp(-i20 x+5.8iz)$ to set the quantum ball in motion with an
initial speed of 18.9 cm/s
and performing real-time simulation with this wave function { in a box of dimension $240\times 240 \times 320 $.
The reflection of the ball from the wall 
is  achieved by  imposing reflecting boundary condition at the surface in the Crank-Nicolson algorithm \cite{CPC}.} 
The uniform motion of the quantum ball without deformation before and after the collision are shown in Figs. \ref{fig9}(a)-(b) and \ref{fig9}(e)-(f), respectively, while its deformation in the proximity of the rigid plane is shown in 
Figs. \ref{fig9}(c)-(d). The quantum ball bounces like an elastic classical  ball 
with the same speed obeying the classical law of reflection.  The results presented 
so far demonstrate beyond doubt that for large velocities the quantum ball interacts elastically with another quantum ball or an external rigid plane with the conservation of kinetic energy. However, the collision is inelastic at small velocities and large deformation of the quantum ball is possible with the nonconservation of kinetic energy.   We performed simulation at small velocities (about one tenth of the velocities considered so far) 
for the interaction with a rigid plane. We find that the quantum ball bounces off the surface at such low incident velocities but with some deformation in shape and nonconservation of kinetic energy resulting in a reduction in the final velocity.

\begin{figure}[!t]

\begin{center}
\includegraphics[trim = 0mm 0mm 0mm 0mm, clip,width=.555\linewidth]{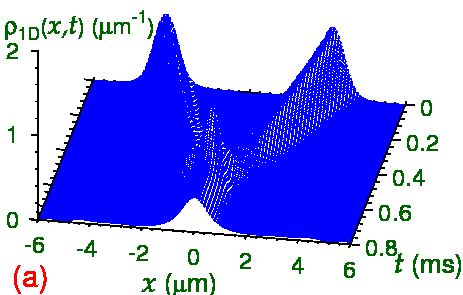}
 \includegraphics[trim = 0mm 0mm 0mm 0mm, clip,width=.43\linewidth]{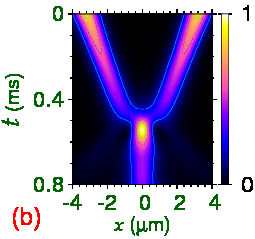}

\caption{ (Color online) 
(a) The 1D density $\rho_{1D}(x,t)$ and (b)
its contour plot during the collision of two  $^7$Li quantum balls, with 
$N=1500$ and $K_3=3\times 10^{-37}(1-0.05i)$ m$^6$/s each,
  initially placed at $x =\pm  3.2$ $\mu$m at $t=0$ and set into motion towards each other 
 with  a velocity 0.45 cm/s, upon 
real-time propagation.  
}\label{fig10} \end{center}

\end{figure}

{To study the inelastic collision dynamics   we consider two  quantum balls each with  $N=1500$ and a three-body term with dissipation, 
 $K_3=3\times 10^{-37}(1-0.05i)$ m$^6$/s,
   place them at $x=\pm 3.2$ $\mu$m, $ y=z=0$ and set them in motion in  opposite directions along the $x$ axis with a small velocity: $v=0.45$ cm/s.
In this case we have used a smaller dissipative term to avoid a large loss of atoms over  long-time collision dynamics  for a small velocity. 
} The dynamics is started   by multiplying the respective 
imaginary-time wave functions by $\exp(\pm i 0.5 x$) 
and  performing real-time simulation for the study of dynamics. { The imaginary and real-time simulations were performed  in boxes of size 
$256\times 256\times 256$ and $512\times 256\times 256$, respectively. }
The dynamics is illustrated by a plot of the time evolution of
1D density $\rho_{1D}(x,t)$ in Fig. \ref{fig10} (a) and the corresponding two-dimensional
contour plot is shown in Fig. \ref{fig10} (b).  
The two quantum balls come close to each other at $x=0$ and
coalesce to form a quantum-ball breather and never separate again. The combined bound 
system remain at rest at $x=0$ 
continuing small breathing oscillation because of a small amount of
 liberated kinetic energy which creates the quantum-ball breather. 
Hence at  sufficiently small incident velocities the collision of two  quantum balls lead 
to the formation of a  quantum-ball breather and at large velocities one has the quasi-elastic collision of two  quantum balls.

\section{Summary}

\label{IV}

We demonstrated the creation of a stable, stationary   BEC  quantum ball (a self-bound BEC) 
under attractive two-body and repulsive three-body contact interactions
  employing a variational and full   numerical solution of the 3D GP equation. The statical properties of the quantum ball are studied by the 
variational approximation and a numerical imaginary-time solution of the 3D GP equation.  
The dynamical properties are studied by a real-time solution of the GP equation including an absorptive three-body term. 
The quantum ball  can move with a constant velocity without deformation.   At large velocities, the collision between the two 
   quantum balls    is   quasi elastic with 
no visible deformation of the final quantum balls.  
We studied head-on and angular collisions of two quantum balls. In all cases, 
unlike classical balls,   the 
balls come out of the collision region without deformation maintaining their  velocities 
and directions of motion unchanged.  In elastic collision of two classical balls one 
can have a change in the direction of their motion subject to energy and momentum conservation.  
At small velocities, the collision between two quantum balls is inelastic with the formation of a quantum-ball breather
after collision.

 { As collapse is not allowed, even in the presence of a very small three-body repulsion, the present suggestion of 
realizing a  trapless BEC quantum ball seems to be attractive from an experimental point of view.
The size of a trapped dipolar BEC is
determined by the harmonic oscillator lengths of the trap, whereas the size of the
present quantum ball is determined by the two-body and three-body interactions. One should start
with a tapped dipolar BEC for $N < N_{\mathrm{crit}}$ where no droplet can be formed, viz. Fig. 
\ref{fig1}(c). Now using the Feshbach resonance technique, one should make the scattering length
 more attractive to enter the  stable domain from unstable domain in Fig. 
\ref{fig1}(c). If the harmonic trap is weak
then initial   size of the trapped BEC should be large, and by increasing the two-body attraction the size
of the quantum ball  could be made much smaller: the sudden change in size will identify the trapless 
quantum ball.
}

\begin{acknowledgments} 
We thank the Funda\c c\~ao de Amparo 
\`a
Pesquisa do Estado de S\~ao Paulo (Brazil)
(Project:  2012/00451-0
  and  the
Conselho Nacional de Desenvolvimento   Cient\'ifico e Tecnol\'ogico (Brazil) (Project: 303280/2014-0) for 
support.
\end{acknowledgments}

%
\end{document}